\documentclass[12pt,a4wide]{amsart}

\usepackage{graphicx}
\usepackage{amsmath}
\usepackage{amssymb}
\usepackage{amsfonts}
\usepackage{subfigure}
\usepackage{yfonts}
\usepackage{setspace}
\usepackage{psfrag}

\usepackage[all]{xy}
\usepackage{epsfig}
\usepackage{amsmath}
\usepackage{amssymb}
\usepackage{amscd}
\usepackage{epsfig}

\newtheorem{lemma}{Lemma}[section]
\newtheorem{theorem}[lemma]{Theorem}
\newtheorem{proposition}[lemma]{Proposition}
\newtheorem{corollary}[lemma]{Corollary}

\DeclareMathAlphabet{\mathpzc}{OT1}{pzc}{m}{it}

 \newcommand{\Q}{{\mathcal Q}}
\newcommand{\eb}{ {\,\,\mathop{\equiv}\limits^{\cl}} \,\,}
\newcommand{\cy}{{\bf\frak c}}

\newcommand{\eu}{ {\,\,\mathop{\equiv}\limits^{\cl_U}} \,\,}

\newcommand{\ew}{ {\,\,\mathop{\equiv}\limits^{\cl_7}} \,\,}
\newcommand{\0}{{\emptyset}}

\newcommand{\rr}{{\mathbb R}}

\newcommand{\cs}{{\mathcal S}}
\newcommand{\rM}{{\mathbb M}}

\newcommand{\cl}{{\mathcal L}}

\newcommand{\ch}{{X \choose 2}}
\newcommand{\E}{{\mathcal E}}

\newcommand{\RR}{{\mathbb R}}

\newcommand{\cP}{{\mathcal P}}

\newcommand{\cC}{{\mathcal C}}

\newcommand{\ra}{{\rightarrow}}
\newcommand{\Ra}{{\Rightarrow}}

\newcommand{\pf}{\noindent{\em Proof: }}
\newcommand{\epf}{\hfill\hbox{\rule{3pt}{6pt}}\\}

\begin{document}

\title{`Lassoing' a phylogenetic tree I: Basic properties, shellings,
and
covers}

\author{
Andreas~W.M.~Dress
\and
Katharina~T.~Huber
\and
Mike~Steel}

\address{
A. W. M. Dress,
CAS-MPG Partner Institute and Key Lab for Computational Biology, Shanghai;
Universit\"at Bielefeld;
Wissenschaftliches Zentrum at infinity$^3$ GmbH (Bielefeld);
and MPI for Mathematics in the Sciences (Leipzig).\\
\email{andreas.dress@infinity-3.de}\\
\and
K. T. Huber,
School of Computing Sciences, University of East Anglia, UK.\\
Tel.: +44-1603-59-3211\\
Fax.: \\
\email{Katharina.Huber@cmp.uea.ac.uk}\\
\and
M. A. Steel,
Department of Mathematics and Statistics, University of Canterbury, New Zealand.\\
Tel.: +64-3-364-2987 ext 7688\\
Fax.: +64-3-364-2587\\
\email{mike.steel@canterbury.ac.nz}\\
}

\date{\today}

\maketitle

\begin{abstract}
A classical result, fundamental to evolutionary biology, states  that an edge-weighted tree $T$ with leaf set $X$, positive edge weights, and no vertices of degree $2$ can be uniquely reconstructed from the 
leaf-to-leaf distances between any two elements of $X$.   In biology,  $X$ corresponds to a set of taxa (e.g. extant species),
the tree $T$ describes their phylogenetic relationships, the edges correspond to earlier species evolving for a time until splitting in
two or more species by some speciation/bifurcation event, and their length corresponds to the genetic change
accumulating over that time in such a species.
In this paper, we investigate which subsets of  $\binom{X}{2}$ suffice
to determine (`lasso')  
the tree $T$
from the leaf-to-leaf distances induced by that tree.  The question is particularly topical since reliable estimates of genetic distance -
even (if not in particular)
by modern mass-sequencing methods  - are, in general, available only for certain combinations of taxa.
\end{abstract}

\section{Introduction}
\label{intro}
A metric $D$ on a finite set $X$ is said to be a `tree metric' if there is a finite tree with leaf set $X$ and non-negative edge weights so that, for all $x,y \in X$, $D(x,y)$ is the path distance in the tree between $x$ and $y$.  It is well known that not every metric is a tree metric. However, when a metric $D$  is a tree metric, the tree (together with its edge weights) that provides a representation of $D$  is -- up to canonical isomorphism -- unique if we also insist that 
the tree is an `edge-weighted $X-$tree', i.e., that
it has no vertices of degree $2$ and that
all of its interior edges have strictly positive edge weights.  However, not all of the $\binom{|X|}{2}$ pairs of distances are required in order to reconstruct the underlying tree.
Thus, it seems of some interest to investigate
which subsets of $\binom{X}{2}$ suffice
to determine (`lasso') the tree.
In this first of a series of papers,
we expound various aspects of this problem, present some relevant definitions, and collect some basic facts.

Our work is partly motivated by the widespread use of distance-based methods for reconstructing phylogenetic  trees in evolutionary biology \cite{fel}. A further reason is that asking similar questions for induced subtrees
rather than for `sparse'  sets of distances gave rise to a rather appealing theory dealing with `sparse' collections of induced subtrees that suffice to 
`define' an $X-$tree (see e.g.~\cite{boc}, \cite{Dre10}).

Provided one has access to all distances,  and these are known to be sufficiently close to the distances induced by some (as yet unknown) tree, then that tree, together with its edge weighting, can be computed -- with some degree of confidence -- from those distances in polynomial time (for example, by using Neighbor-Joining \cite{att}). However, much of the data being generated -- even by modern genomic methods -- have patchy taxon
coverage \cite{phil} whereby only certain pairs of taxa have a known (or, at least, sufficiently reliable) distance.  This raises interesting mathematical questions (besides the obvious  statistical and algorithmic ones) concerning tree reconstruction from such incomplete data some of which we will address here.

More specifically, in this first of a series of papers,
we want to explore the basic properties of `edge-weight', `topological', and `strong lassos' -- being primarily interested in the uniqueness question: Given the restriction of a tree metric $D$ to some subset $\cl$ of $\binom{X}{2}$,
how much can we learn about the tree representing $D$ from that restriction?
In particular, we ask which subsets  $\cl$ of $\ch$ provide enough `coverage' in order to fully determine an edge-weighted $X-$tree or, at least, its shape,  or -- given its shape -- its edge lengths in terms of just
the distances it induces between the pairs of taxa collected in $\cl$. Or, put differently, how much `missing data' (pairs of taxa $x,y$ for which $D(x,y)$ is not known) can we allow
and still be guaranteed to recover them from those distances that we can observe.

\section{Some basic definitions and facts}
\label{definitions}

\subsection{Trees and tree metrics}
\label{ttm}

Consider any finite tree $T=(V,E)$ with vertex set $V$, leaf set $X\subseteq V$, and edge set $E\subseteq \binom{V}{2}$ together with an {\em edge weighting} -- i.e.,  a map  $\omega$ in the set $\Omega = \Omega_T:=\rr_{\ge 0}^E$ that assigns a non-negative length $\omega(e)$ to every
edge $e \in E$.  Any such pair $(T, \omega)$  induces a distance function:
\begin{equation}
D_\omega=D_{(T,\omega)}: \ch \rightarrow \RR_{\ge 0}: \{x,y\}\mapsto D_\omega(x,y) :=
\omega_+\big(E_T(x|y)\big)
\end{equation}
where $E_T(u|v)$ denotes, for any two vertices $u,v\in V$, the set of edges  in $E$ that `separate' $u$ and $v$ in $T$ (and, thus, together make up the path from $u$ to $v$ in $T$) and $\omega_+(F)$ denotes, for any non-empty subset $F$ of $E$, the sum $\sum_{e\in F}\omega(e)$.

\begin{figure}[ht]
\begin{center}
\resizebox{4cm}{!}{
\includegraphics{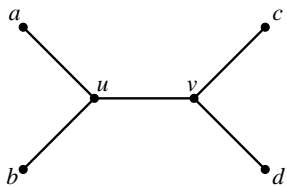}
}
\caption{ 
The set  $\cl_4 =\{\{a,b\}, \{c,d\},\{ a,c\},\{ b,d\}\}$ lassos the shape 
of the $X_4-$tree $T_4$
while $\cl_4 \cup \{\{a,d\}\}$
and $\cl_4 \cup \{\{b,c\}\}$ are strong lassos for $T_4$ (see text for details).}
\label{figure_quartet}
\end{center}
\end{figure}

For example,  in Fig. \ref{figure_quartet}, we have
$E_T(a|c)=\big\{\{a,u\},\{u,v\},\{v,c\}\big\}$ and, thus,
$D_\omega(a,c) = 3$ for the
binary tree $T:=T_4$ with leaf set $X_4:=\{a,b,c,d\}$ and an interior edge that separates the leaves $a,b$ from the leaves $c,d$
provided unit edge length has been assigned to
all edges of that tree.

\smallskip
\noindent
While $D_\omega$ is clearly a (pseudo-)metric on $X$
(and a proper metric if -- but not necessarily only if --
$\omega$ is strictly positive),
not every metric on $X$ can be represented in this way: The condition for an arbitrary metric $D$ on $X$ to have a {\em
phylogenetic representation}, that is, to be representable in the form
$D=D_\omega$ for some finite edge-weighted tree $(T,\omega)$ with leaf set $X$, is that $D$ satisfies
 the well-known {\em four-point condition} which states that, for all $a,b,c,d\in X$, the larger two of the three distance sums $D(a,b)+ D(c,d), \, D(a,c) + D(b,d), \,D(a,d) + D(b,c)$ coincide or, equivalently, if
 \begin{equation}
 \label{D(|)}
 D(ab|cd):=\max\big\{ D(a,c) + D(b,d), \,D(a,d) + D(b,c)\big\}-D(a,b)-D(c,d)
\end{equation}
 is non-negative for all  $a,b,c,d\in X$.

 Such a metric $D$ is said to be a {\em tree metric}, and any finite tree $T=(V,E)$ as above for which some
$\omega\in \Omega_T$ with $D=D_\omega$ exists will be dubbed a {\em $D-$tree}. Furthermore, such a tree $T$ will be said to be a {\em proper $D-$tree} if $T$ has no vertices of degree $2$ and it has a {\em proper} edge weighting $\omega$ with $D=D_\omega$, i.e., a map $\omega\in \Omega_T$ that is strictly positive on all interior edges of $T$.

Clearly, given any tree metric $D$,  many non-equivalent $D-$trees $T$ with edge weightings $\omega$ can exist such that $D=D_\omega$ holds, since adding zero-length edges and/or subdividing any edge of a $D-$tree by degree $2$ vertices yields further $D-$trees.
However, it has been well known since the $1960$s (see, for instance,~\cite{bar} and \cite{sem} and the references therein) that there is `essentially' only one proper $D-$tree $T$ for any tree metric $D$ and, given $T$,
 only one edge weighting $\omega\in \Omega_T$ for which $D=D_\omega$ holds.  This was actually one of the starting points of what currently is called {\em phylogenetic combinatorics}.

More specifically, recall that, given a finite set $X$ of cardinality at least
$3$
(the set $X$ typically represents the collection of `taxa' under consideration -- e.g. some
extant species),  a finite tree $T=(V,E)$ with vertex set $V$, leaf set $X\subseteq V$, and edge set $E\subseteq \binom{V}{2}$ having no vertices of degree $2$ is said to be a {\em phylogenetic $X-$tree} or (in the context of this paper) more briefly an {\em $X-$tree} 
and that an
$X-$tree for which every interior vertex has degree $3$ is said to be a {\em binary $X-$tree}.  With these definitions in hand, the following relationships are well-known and easily established. 

\begin{itemize}
\item[(i)]
$|E|\le 2|X|-3$ holds for every $X-$tree $T=(V,E)$; and
\item[(ii)]
$|E|= 2|X|-3$ holds if and only if $T$ is a binary $X-$tree 
\end{itemize}
Recall also that two $X-$trees $T=(V,E)$ and $T'=(V', E')$ are said to be ({\em topologically}) \!{\em equivalent}  (written $T \simeq T'$) if there exists a (necessarily unique)
graph isomorphism $\varphi: T  \mbox{ } \tilde\ra  \mbox{ } T'$ that respects $X$, i.e.,  a bijection $\varphi: V   \mbox{ } \tilde\ra  \mbox{ } V'$ with
$E'=\big\{\{\varphi(u),\varphi(v)\}:\{u,v\}\in E\}$
and $\varphi (x) = x$
for all $x \in  X$, and that $T'$ is defined to be a  {\em refinement} of $T$ (written $T \le T'$) if -- up to equivalence -- $T$ can be obtained from $T'$ by collapsing edges in $T'$ (see \cite{sem}).
Furthermore,  two edge-weighted
 $X-$trees
$(T,\omega),  (T', \omega')$ are said to be {\em isometric}
$\big(\mbox{written  }  (T,  \omega) \equiv (T', \omega')\big)$ if there exists a graph isomorphism $\varphi: T  \mbox{ } \tilde\ra  \mbox{ }  T'$ as above that respects not only $X$, but also the edge lengths, i.e.,  also $\omega(\{u,v)\})=\omega'(\{\varphi(u),\varphi(v)\})$ holds for all edges $\{u,v\}\in E$ of $T$.  For example, denoting the
`all-one map'
on a set $A$ by ${\bf 1}^A$, two $X-$trees $T=(V,E)$ and $T'=(V'E')$  are equivalent if and only if the corresponding edge-weighted
 $X-$trees
$(T,{\bf 1}^E)$ and $(T', {\bf 1}^{E'})$ are isometric.

The
basic
result referred to above then states that, given any 
two $X-$trees  $T,T'$ with proper edge weightings 
$\omega\in \Omega_T$ and $\omega'\in \Omega_{T'}$, one has
\begin{equation}\label{basic'}
D_\omega=D_{\omega'}\iff (T,\omega) \equiv(T',\omega')
\end{equation}
and, therefore, also
\begin{equation}\label{basic}
D_\omega=D_{\omega'}\iff \omega=\omega'
\end{equation}
for any fixed $X-$tree $T=(V,E)$ and all $\omega,\omega' \in \Omega_T$.

\medskip
What we will be concerned with here is that, given $(T,\omega)$ and $(T',\omega')$ as above, we do not even always need the associated metrics $D_\omega$ and $D_{\omega'}$ to coincide on {\bf\em all} pairs $\{x,y\}\in \ch$ to conclude

-- that $T'$ must be equivalent to (or at least a refinement of) $T$,

-- that $(T,\omega)$ and $(T',\omega')$
must be isometric, or

-- that $\omega=\omega'$ must hold
in case $T=T'$.

\noindent
Indeed, if 
$T$ and $T'$ are
two 
$X_4-$trees, 
and $\omega$ and $\omega'$ are proper edge weightings of $T$ and $T'$, respectively, then

(i)  $T$ and $T'$ must be equivalent whenever the
 two metrics $D:=D_\omega$ and $D':=D_{\omega'}$ coincide on
the four pairs $\{a,b\}, \{c,d\}, \{a,c\},$ and $\{b,d\}$, and
$D(a,b)+D(c,d)<
D(a,c)+D(b,d)$ holds (in which case, both must be
equivalent to the tree depicted in Fig.~\ref{figure_quartet});

(ii)
$(T,\omega)$ and $(T',\omega')$ must be isometric
or, equivalently, $D$ and $D'$ must coincide
if these two maps coincide, in addition, on just one of the remaining two 
pairs $\{a,d\}$ or $\{b,c\}$.

\subsection{Lassos}
\label{lassos}

To deal with such matters, we define, given a subset $\cl$ of $\ch$, two edge 
weighted $X-$trees $(T,\omega)$ and $(T',\omega')$ to be {\em $\cl$-isometric} 
 \big(written
$(T,\omega)\eb(T',\omega')$\big)  if $D|_\cl=D'|_\cl$ holds for $D:=D_\omega$ 
and $D':=D_{\omega'}$.
Then, given an $X-$tree $T$, it seems  of some interest to study those subsets $\cl$  
of $\binom{X}{2}$ that have one of the following properties:
\begin{itemize}
\item[(L-i)] $\omega = \omega'$ holds for
all proper edge weightings $\omega, \omega'$ of $T$ with $(T,\omega)\eb(T,\omega')$;
  \item[(L-ii)] $T \simeq T'$ holds for any $X-$tree $T'$ for which there exist 
proper edge weightings
$\omega$ of $T$ and $\omega'$ of $T'$
with $(T,\omega)\eb(T',\omega')$;
 \item[(L-ii$'$)] $T \le T'$ holds for any $X-$tree $T'$ for which there exist proper edge weightings
$\omega$ of $T$ and $\omega'$ of $T'$
with $(T,\omega)\eb(T',\omega')$;
\item[(L-iii)] $(T,\omega) \equiv (T',\omega')$ holds,
for  every given proper edge weighting $\omega$ of $T$, for any
$X-$tree $T'$ and any proper edge weighting $\omega'$ of $T'$ with $(T,\omega)\eb(T',\omega')$.
\end{itemize}

\noindent
To this end, given an $X-$tree $T$, we define  a subset $\cl$  of $\binom{X}{2}$ to be:
\begin{itemize}
\item[(i)]  an {\em edge-weight lasso for $T$ } (or {\em to lasso the edge weights of $T$}) if
(L-i) holds;
\item[(ii)]   a {\em topological lasso for $T$ }  (or {\em to lasso the shape of  $T$}) if (L-ii) holds;

\item[(ii$'$)]   a {\em weak lasso for $T$ }  (or {\em to corall $T$}) if (L-ii$'$) holds;  and

\item[(iii)]   a {\em strong lasso for $T$ } (or just {\em to lasso $T$})
whenever (L-iii) holds.
\end{itemize}

As we deal  here with `lassos', any $2$-subset $\cy=\{x,y\}\in \ch$ of $X$
will also be called a {\em cord}, often written more briefly as $\cy=xy$;
also, we refer to the cords in a lasso $\cl$ as the cords `in' $\cl$.

Using this terminology, we can rephrase the example discussed at the end of Subsection \ref{ttm} as follows: The four cords $ab$, $cd$, $ac$ and $bd$ form a topological lasso $\cl_4$ for the tree $T_4$
depicted in Fig. \ref{figure_quartet}, and adding either the cord  $ad$ or $bc$ yields a strong lasso for that tree.

Clearly, a subset $\cl$ is an edge-weight lasso
for an $X-$tree $T=(V,E)$ if and only if  $D=D'$ holds for any two tree metrics
$D,D'$ defined on $X$ with $D|_\cl=D'|_\cl$ for which $T$ is simultaneously a proper $D$- and a proper $D'-$tree: Indeed, if  $\omega$ and
$\omega'$ are proper edge weightings of  $T$ with
$D=D_\omega$ and $D'=D_{\omega'}$, we have ``$D|_\cl=D'|_\cl\iff (T,\omega)\eb (T,\omega')$'' and ``$D=D'\iff \omega=\omega'$'' and, therefore, ``$D|_\cl=D'|_\cl\,\Ra\, D=D'$'' if and only if
``$(T,\omega)\eb (T,\omega')\, \Ra \, \omega=\omega'$''.

Similarly, $\cl$ is a topological (or a weak) lasso for $T$ if and only if every $X-$tree $T'$ for which there exist tree metrics $D$ and $D'$  with $D|_\cl=D'|_\cl$ such that
$T$ is a proper $D-$tree and $T'$ is a proper $D'-$tree is
 equivalent to (or a refinement of) $T$.
And $\cl$ is a strong lasso for $T$
if and only if $D=D'$ holds for any two tree metrics
$D,D'$ defined on $X$ with $D|_\cl=D'|_\cl$ for which $T$ is a proper $D-$tree and, hence,
if and only if it is both,
an edge-weight lasso
and a topological lasso for $T$.

 In particular, if
there exists a pair $\omega, \omega'$ of edge weightings of $T$ with $D_\omega|_\cl =D_{\omega'}|_\cl$ such that
$\omega$ is a proper and $\omega'$ is not a proper edge weighting of $T$, then $\cl$ is neither a topological lasso for $T$ nor for the $X-$tree that results by `collapsing' any of the interior edges $e$ of $T$ with $\omega'(e)=0$.

\medskip
\subsection{Some further conventions, definitions,
notations, and well-known facts}
\label{notation}
We end this section by listing some simple conventions,
definitions,
and well-known facts (see, e.g., \cite{sem}) that will be used throughout.

\smallskip
\noindent
{\bf 2.3.1\,\,} Firstly, we will assume throughout that $X$ is a finite set of cardinality
$n \geq 3$ and we put $\bigcup \cl:=\bigcup_{\cy\in \cl}\cy$ for any non-empty subset $\cl\subseteq \ch$.
We will refer to a subset $\cl$ of $\ch$ as being `connected', `disconnected' or `bipartite' etc. whenever the graph $\Gamma(\cl):=(X,\cl)$ is connected, disconnected, or bipartite and so on, and a connected component of $\Gamma(\cl)$ will also be called a
connected component of $\cl$.

\smallskip
\noindent
{\bf 2.3.2\,\,}For every edge $f$ of a tree $T=(V,E)$,
we denote by $\delta_f \in \Omega_T$ the map defined by
\begin{equation}
\label{eeq}
\delta_f: E\ra \rr: e\mapsto
\delta_{e,f}:=
\begin{cases}
      & 1, \text{ if } e=f; \\
      & 0, \text{ otherwise.}
\end{cases}
\end{equation}
And for every leaf $a$ of 
a tree $T$ 
with at least $2$ vertices, 
we denote by $e_a=e_a^T$ the unique
edge of $T$ containing $a$ and by $v_a$ the other
(in case $|V|\ge 3$ necessarily interior)
vertex of $T$
contained in $e_a$.

\smallskip
\noindent
{\bf 2.3.3\,\,}Two distinct leaves,  $a$ and $b$, in a tree $T$ with $v_a = v_b$ will
be said to form a 
{\em $T-$cherry}, 
and they will be said to form a 
{\em $T-$proper cherry}
if, in addition,  $v_a (=v_b)$ has degree $3$; for example, the two pairs $a,b$ and $c,d$ form proper cherries in the tree
$T_4$
depicted in Fig.~\ref{figure_quartet}.
A {\em caterpillar tree} 
is a binary $X-$tree that has exactly two proper cherries (see, for example the tree
$T_6$ in Fig.~\ref{figure_caterpillar}, or the tree in Fig. ~\ref{caterfig}).

\smallskip
\noindent
{\bf 2.3.4\,\,}The {\em median} of three vertices $u,v,$ and $w$ of a tree $T=(V,E)$ is the unique vertex in $V$ that is simultaneously contained in the three paths connecting any two of $u,v,$ and $w$ in $T$, and will be denoted by ${\rm med}_T(u,v,w)$. For example, the vertex $u$
in Fig.~\ref{figure_quartet}  is the median of the three leaves $a,b,c$.

\begin{figure}[ht]
\begin{center}
\resizebox{12cm}{!}{
\includegraphics{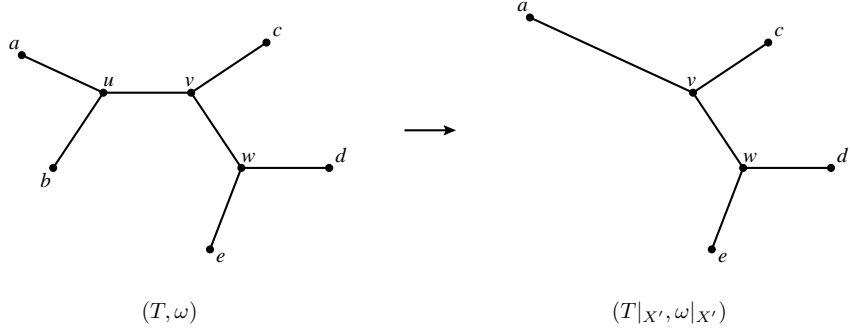}
}
\caption{
For $X':=\{a,c,d,e\}\subset X_5=\{a,b,c,d,e\}$,
the $X'-$tree on the right is obtained from the
$X_5-$tree $T_5$ 
on the left by restricting its leaf set to 
$X'$. 
The associated induced edge weighting $\omega_{|X'}$ is also indicated.}
\label{fig:induced}
\end{center}
\end{figure}

Given an $X-$tree $T=(V,E)$ and any subset $X'$ of $X$, the {\em restriction} of $T$ to
$X'$ (i.e.,  the tree with vertex set ${\rm med}_T(X'):=\{{\rm med}_T(x,y,z):x,y,z\in X'\}$ and edge set the set of all pairs $\{u,v\}\in {{\rm med}_T(X') \choose 2}$ for which ${\rm med}_T(u,v,x)\in \{u,v\}$ holds for all $x\in X'$) will be denoted by $T|_{X'}$, and its vertex and edge
sets
by $V|_{X'}$ and $E|_{X'}$, respectively. And, given any edge weighting $\omega$ of $T$, the {\em induced}  edge weighting of $T'$, i.e.,   the edge weighting that maps any edge 
$\{u,v\}\in E|_{X'}$ onto the sum $\omega_+\big(E_T(u|v)\big)$, will also be denoted by $\omega|_{X'}$.  These concepts are illustrated in Fig. \ref{fig:induced}
for the caterpillar tree $T_5$ with leaf set 
$X_5:=\{a,b,c,d,e\}$ 
and the two cherries $a,b$ and $d,e$
depicted in Fig.\ref{fig:induced} on the left.

It is well known and easily seen that $T|_{X'}$ is a (binary) $X'-$tree for every (binary) $X-$tree $T$ and every subset $X'$ of $X$ of cardinality at least $3$.

\smallskip
\noindent
{\bf 2.3.5\,\,}
An {\em $X-$split} is a `split' or `bipartition' of $X$ into two disjoint non-empty subsets.
A {\em quartet} is  a bipartition
of a $4$-set into two disjoint subsets of cardinality $2$.
In case $a,a',b,b'$ are any $4$ distinct elements, the quartet $\big\{\{a,a'\},\{b,b'\}\big\}$
is also denoted, for short, by $aa'\|bb'$ while $a|a'|b|b'$ stands for the partition of $\{a,a',b,b'\}$ into the four one-element sets $\{a\},\{a'\},\{b\},\{b'\}$.

 A {\em quartet tree}  is a binary tree
 $T$ with exactly four leaves  -- and, therefore, exactly two cherries.
 We will also say that such a tree $T$ is a {\em quartet tree of type} $aa'\|bb'$ if its two cherries are formed by the leaves $a,a'$ and $b,b'$, i.e.,  $\{a,a',b,b'\}$ is  
 the 
 $4$-set that forms the leaf set of $T$, and $T$  has a (necessarily interior and necessarily unique) edge
 that separates $a,a'$ from $b,b'$ (so, as stated in Fig.\,\ref{figure_quartet},
the tree $T_4$ depicted in that figure is a quartet tree of type $ab\|cd$). In addition, a tree $T$ with exactly four leaves $a,a',b,b'$ will be said to be a tree of type $a|a'|b|b'$ if it is non-binary, so that any tree $T$ with leaf set $\{a,a',b,b'\}$ is either a tree of type $a|a'|b|b'$ or a quartet tree of type $aa'\|bb'$,  $ab\|a'b'$, or  $ab'\|a'b$.

Further, an $X-$tree $T$ is said to {\em display}
a quartet $xx'\|yy'$ (or, respectively, the partition $x|x'|y|y'$) if $\{x,x',y,y'\}$ is a $4$-subset of $X$ and
$T|_{\{x,x',y,y'\}}$ is a quartet tree of type $xx'\|yy'$ (or, respectively,  a tree of type $x|x'|y|y'$).  By abuse of notation,
$T$ will also be said to display $xx'|yy'$ if it either displays $xx'\|yy'$ or $x|x'|y|y'$ or, equivalently, neither  $xy\|x'y'$ nor $xy'\|x'y$.
The collection of all quartets displayed by $T$ will be denoted by $\Q(T)$.

Recall also that, given any five distinct elements $x,x',y,y',y''\in X$, $T$ displays $xx'\|yy''$ (or $xx'|yy''$, respectively) if it displays  $xx'\|yy'$ and $xx'\|y'y''$ (or $xx'|yy'$ and $xx'|y'y''$) \cite{col}. In addition, given any proper edge weighting $\omega$ of $T$,
$T$ displays
\begin{itemize}
\item
$xx'\|yy'$ if and only if
$D_\omega(xx'|yy')>0$ holds\footnote{Recall that $D_\omega(ab|cd) = \max\big\{ D(a,c) + D(b,d), \,D(a,d) + D(b,c)\big\}-D(a,b)-D(c,d)$ for $D=D_\omega$},
\item  $x|x'|y|y'$ if and only if  $D_\omega(x,y)+D_\omega(x',y')=D_\omega(x,y')+D_\omega(x',y)=D_\omega(x,x')+D_\omega(y,y')$ holds; and
\item  $xx'|yy'$ if and only if  $D_\omega(x,y)+D_\omega(x',y')=D_\omega(x,y')+D_\omega(x',y)$ holds.
\end{itemize}
Furthermore (see for instance Chapter 7 in \cite{Dre10}), one has
 \begin{equation}\label{rho}
\min\big\{D_\omega( xx'|yy'),D_\omega( xx'|y'y'')\big\}\le D_\omega( xx'|yy'')
\end{equation}
for all $x,x',y,y',y''$ as above whenever $T$ displays $xx'|yy'$ and $xx'|y'y''$.

In consequence, 
\begin{equation}\label{=}
D_\omega( xx'|yz)= D_\omega( xx'|y'z)
\end{equation}
holds for all  $x,x', y,y',z\in X$ with 
$D_\omega( xx'|yy')> D_\omega( xx'|yz)$ and, given any six elements $x,x',y,y',$ and $z,z'$, one has
\begin{equation}\label{==}
D_\omega( xx'|yz)=D_\omega( xx'|y'z)=D_\omega( xx'|yz')=D_\omega( xx'|y'z')
\end{equation}
whenever
$D_\omega( xx'|yy'),D_\omega( xx'|zz')> D_\omega( xx'|yz)$ holds.

\medskip
\noindent
{\bf 2.3.6\,\,}
Next, given any two non-empty subsets $A$ and $B$ of $X$, an $X-$tree
$T$ is said to {\em display} $A\|B$ (or $A|B$, respectively), if $A$ and $B$ are disjoint and $T$ displays $aa'\|bb'$ (or $aa'|bb'$, respectively) for any two distinct elements $a,a'\in A$ and $b,b'\in B$ or, equivalently, if this holds for some fixed $a\in A$ and $b\in B$ and all $a'\in A-\{a\}$ and  $b'\in B -\{b\}$.

If $T$ displays $A\|B$ and $A\cup B=X$ holds, the pair $A,B$ will also be called a  {\em  $T-$split}, and a {\em non-trivial} $T-$split if, in addition,
$|A|,|B|\ge 2$ holds. Similarly, if $T$ displays $A|B$ and $A\cup B=X$ holds, the pair $A,B$ will also be referred to as a {\em virtual  $T-$split}, and a {\em non-trivial} virtual $T-$split if, in addition,
$|A|,|B|\ge 2$ holds.

Notice that if $T$ displays both $A\|B$ and $A'\|B'$ (or $A\|B$ and  $A'|B'$ or $A|B$ and $A'\|B'$) then one of the four intersections $A\cap A', A\cap B', B\cap A'$, and $B\cap B'$ is empty. Any two $X-$splits that satisfy this last property are said to be {\em compatible}, otherwise they are {\em incompatible}.

Further, given -- in addition -- any edge weighting $\omega$ of $T$, we put
$$
D_\omega(A|B):=\min\big\{D_\omega(aa'|bb'):a,a'\in A, b,b'\in B\big\}
$$
so that $T$ displays $A\|B$ if and only if $D_\omega(A|B)>0$ holds for one or, equivalently, for every proper edge weighting $\omega$ of $T$ --
note that this notation is consistent with our previous notation as, in view of the triangle inequality, we have $D_\omega(aa'|bb')=D_\omega(\{a,a'\}|\{b,b'\})$ for all $a,a',b,b'$ in $X$.

\smallskip

Clearly, two leaves $a,a'\in X$ form a proper $T-$cherry
if and only if 
the pair $\{a ,a'\},X-\{a,a'\}$ forms a $T-$split or, equivalently, if and only if 
$T$ displays $\{a ,a'\}\|X-\{a,a'\}$;  and they form just a $T-$cherry if and only if the pair $\{a ,a'\},X-\{a,a'\}$ forms a virtual $T-$split or, equivalently, if and only if $T$ displays $\{a ,a'\}|X-\{a,a'\}$. So, both trees depicted in
Fig.\,\ref{fig:induced} display the quartet $ac\|de$;
and the pair $\{a,b,c\}, \{d,e\}$ forms a $T_5-$split.

\smallskip
\noindent
{\bf 2.3.7\,\,}
It is also well known that an $X-$tree $T$ displays $A\|B$ for two disjoint non-empty subsets $A$ and $B$ of $X$ if and only if there exists some edge $e\in E$ with $e\in E_T(a|b)$ for all $a\in A$ and $b\in B$ and, hence, if and only if there exists a $T-$split
$A^*, B^*$ of $X$ with $A\subseteq A^*$ and $B\subseteq  B^*$. Furthermore, if $A,B$ is a $T-$split, there is exactly one edge $e=e_{A\|B}\in E$ with $e\in E_T(a|b)$ for all $a\in A$ and $b\in B$. And associating, to each $T-$split $A,B$, the edge $e_{A\|B}$ defines a canonical one-to-one correspondence between the collection $\cs(T)$ of all $T-$splits and the edge set $E$ of $T$ as well as between the collection $\cs_{nt}(T)$ of all non-trivial $T-$splits and the set of all interior edges of $T$.

Furthermore, given any bipartition $S'$ of $X$ into two disjoint and non-empty subsets  $A',B'$ of $X$ and any $X-$tree $T$, the following assertions are equivalent:
\begin{itemize}
\item
the pair $A',B'$ forms a virtual $T-$split;
\item
$S'=\{A',B'\}$ is  compatible with every $T-$split $S\in \cs(T)$;
\item
there exists an $X-$tree $T'$ with $\cs(T')=\cs(T)\cup \{S'\}$ such that collapsing the edge $e_{A'\|B'}$ in $T'$ yields -- up to canonical isomorphism -- the tree $T$.
\end{itemize}
\smallskip

\begin{figure}[h]
\begin{center}
\resizebox{13cm}{!}{
{\includegraphics{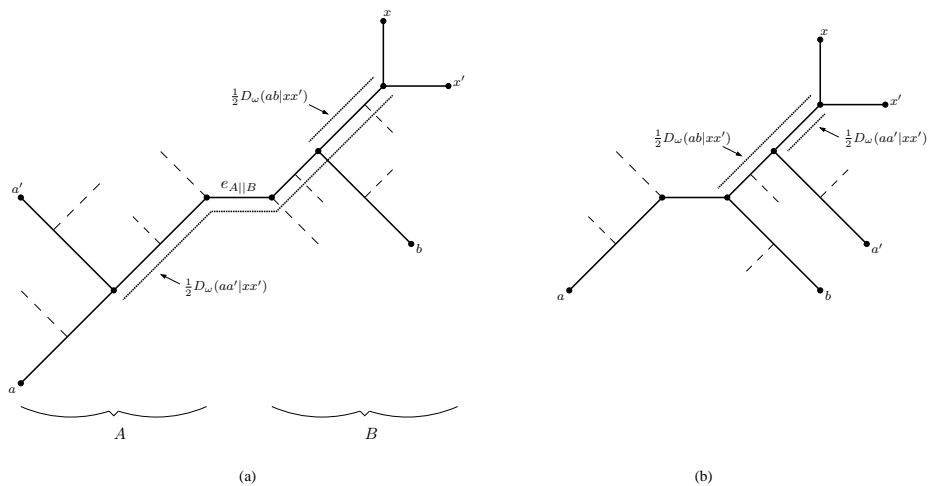}}
}
\caption{(a) When $A,B$ forms a $T-$split  and $x,x' \in B$, then $D_\omega(aa'|xx')>D_\omega(ab|xx')$ holds for all 
$a,a'\in A$ 
and $b\in B$; (b) If $A,B$ does not form a $T-$split, there  
exists, for all $a\in A$ and $x,x'\in B$, some $a' \in A$ and $b \in B$ 
with $D_\omega(aa'|xx')\leq D_\omega(ab|xx')$ (see text for details).}
\label{new_fig}
\end{center}
\end{figure}

And putting $A(a|bx):=\{a'\in X:a'=a \text{ or }aa'\|bx\in \Q(T)\}$ for any three distinct elements $a,b,x\in X$, the pair $A(a|bx), X-A(a|bx)$ always forms a $T-$split and the pair $A(a|bx)\cup A(b|ax), X-\big(A(a|bx)\cup A(b|ax)\big)$ forms a virtual $T-$split. Furthermore, there exist, for every $T-$split $S=A|B$ with $|B|>1$, two distinct elements $b,x\in B$ such that $S$ coincides with the pair $A(a|bx), X-A(a|bx)$ for one or, equivalently, for every $a\in A$. Also, given any element $x'\in X-\{a,b,x\}$ with $ab\|xx'\in \Q(T)$, one has $a'\in A(a|bx)$ for some $a'\in X-\{a,b,x\}$ if and only if one has $D_\omega(aa'|xx')>D_\omega(ab|xx')$ for some or, equivalently, for every proper edge weighting $\omega$ of $T$. In particular, given any bipartition $S$ of $X$ into two disjoint and non-empty subsets  $A,B$ of $X$, some $a\in A$ and two distinct elements $x,x'\in B$, the pair $A,B$ forms a
$T-$split if and only if one has $D_\omega(aa'|xx')>D_\omega(ab|xx')$ for all $a'\in A$ and $b\in B$ (see Fig. ~\ref{new_fig}).

\smallskip
\noindent
{\bf 2.3.8\,\,}
Finally, it is also well known (see e.g. \cite{sem}) that, given
two $X-$trees $T$ and $T'$, one has
$$
T\simeq T'\iff \Q(T)=\Q(T')\iff \cs(T)=\cs(T') \iff \cs_{nt}(T)=\cs_{nt}(T')
$$
or, more generally,
$$
T\le T'\iff \Q(T)\subseteq \Q(T')\iff \cs(T) \subseteq \cs(T') \iff \cs_{nt}(T) \subseteq \cs_{nt}(T').
$$

\section{Contents and outlook}\label{contents}

Our series of papers devoted to a rather detailed study of edge-weight, topological, weak, and strong lassos
is organized as follows: In the next section (Section~\ref{facts}), we will present
some elementary properties and some instructive examples
of lasso sets. In Section \ref{rec}, we will present and apply some results that are helpful for investigating lassos in a recursive fashion. In  Section \ref{shell}, we will introduce and discuss a useful concept for recognizing strong lassos -- the concept of `$\cl$-shellability',
 -- and, finally, we will study
two particular types of lassos called
{\em $e-$covers} and {\em $t-$covers}, respectively,
in Section \ref{covsec} -- lassos that show up naturally in our context and have, to some extent, already been recognized in previous work (cf. \cite{bar} and \cite{cha}) as exhibiting some particularly attractive and useful properties.

In particular, for any bipartition $A,B$ of $X$, the set 
$\cl = A\vee B:=\big\{\{a,b\}: a\in A, b\in B\big\}$  is a topological lasso
for an $X-$tree $T$ if and only if 
$ A\vee B$ is a $t-$cover of $T$ if and only if 
$A,B$ is incompatible with every non-trivial
virtual $T-$split (Theorem \ref{vee}).

In a subsequent paper, we will discuss various classes of examples and
`counter-examples'. 
In particular, we will present a full characterization of topological
lassos for $X-$trees with at most two interior vertices, we will show
that every $t-$cover of such an $X-$tree $T$ is a weak lasso for $T$
provided both of its two interior vertices have degree at least $4$,
and we will list all edge-weight and all topological lassos for $X-$trees with at most five leaves. That paper will also show, in particular, that all
edge-weight lassos for such a tree are strong lassos and that there are
minimal topological lassos for
every
binary $X-$tree $(V,E)$ with exactly five
leaves (there is `essentially' only one such tree) that have
cardinality $|E|=7$ while most
such lassos are bipartite and have
cardinality $6$.
And, using our recursive approach, we will draw some consequences that are of general interest for lassos for arbitrary $X-$trees.

In addition,
noting that the minimal edge-weight lassos for an $X-$tree $T$ form the set of bases of a certain matroid with point set $\ch$ denoted by $\rM(T)$, we
will study
this matroid in yet another paper.    In particular, we will show that $T$ is determined, up to equivalence, by $\rM(T)$, i.e.,  ``$T \simeq T' \iff \rM(T)=\rM(T')$'' holds for any two $X-$trees $T,T'$. We will also show that
\begin{itemize}
\item[(i)]
 a binary $X-$tree $T$ is a caterpillar tree if and only if the matroid $\rM(T)$ is a binary matroid,
\item[(ii)] a subset $\cl$ of $\ch$ is a strong lasso for some $X-$tree $T$ if and only if it is a non-bipartite topological lasso for $T$ -- more generally, the
co-rank of some connected subset $\cl$ of $\ch$ in $\rM(T)$ that is a weak lasso for $T$ never exceeds $1$, and it coincides with $1$ if and only if $\cl$ is bipartite which can happen for $T$ only if every 
$T-$cherry is a proper $T-$cherry, and
 \item[(iii)]
the edge set $\cl$ of 
a
complete bipartite graph 
with vertex set $X$ is a topological lasso if and only if $\cl$ has co-rank $1$
in $\rM(T)$.
\end{itemize}

We will not deal here with the corresponding `existence question':  Given a subset $\cl$ of $\binom{X}{2}$ and some map $D: \cl \rightarrow X$, when does $D$ extend to a tree metric on $X$?  The computational
complexity of this existence question has been settled,  as it is nothing but  the `Matrix Completion to Additive' problem that -- not unexpectedly -- was shown to be NP-complete (\cite{far}, Theorem 6), and algorithmic approaches to special instances of this problem have already been explored in \cite{gue2}, \cite{gue}, and \cite{will}.

Also, our focus here is on the mathematical, rather than the algorithmic, aspects of the uniqueness question, as the mathematical structure underlying that question appears to be intricate enough already compared with
the case settled long ago in which all distances are known. As such, it seems to deserve  especially dedicated attention.

\section{Some basic properties and some  instructive  examples of lassos}
\label{facts}

Assume throughout this section that $T=(V,E)$ is an $X-$tree and that $\cl$ is a subset  of $\binom{X}{2}$.
Recall that we will often write $xy$ as  a shorthand for $\{x,y\}$.

We begin by noting that edge-weight lassos can be characterized in terms of the linear forms their cords induce
on the real vector space $\rr^E$:
\begin{theorem}\label{LA}
The set $\cl$ is an edge-weight lasso for $T$ if and only if $X=\bigcup\cl$ and there is
 no non-zero map $\omega_0\in \rr^E$ such that the linear maps
\begin{equation}\label{lambdaeq}
\lambda^T_{xy}:  \rr^E\ra \rr: \omega\mapsto
\omega_+\big(E_T(x|y)\big) \,\,\big(= D_{\omega}(x,y)\big) \,
\,\,\,\,\,(xy\in \ch)
\end{equation}
vanish on $\omega_0$ for all $xy\in \cl$.

In particular,
$|\cl| \geq |E|$ must hold for every edge-weight lasso $\cl$ for $T$, and $|\cl| = |E|$ must hold  for every minimal edge-weight lasso $\cl$ for $T$.
\end{theorem}
\pf If $(T,\omega)\eb (T,\omega')$ would hold for two distinct proper edge weightings $\omega,\omega'\in \Omega_T$, we would have $\lambda^T_{xy}(\omega_0)=0$ for all $xy \in \cl$ for the map
$\omega_0:=\omega-\omega'$. And if, conversely,
 $\lambda^T_{xy}(\omega_0)=0$ holds, for all $xy \in \cl$, for some  non-zero map $\omega_0\in \rr^E$,
adding a sufficiently small multiple of $
\omega_0$ to any proper  edge weighting $\omega$ of $T$ would yield a
proper edge weighting $\omega'\neq \omega$ of $T$ with $(T,\omega)
\eb(T,\omega')$.
The last claim follows by applying some basic linear algebra to the bilinear pairing

\medskip
$
\langle ... |...\rangle_T:\rr^\cl\times \rr^E\ra \rr: (\rho,\omega)\mapsto\langle \rho|\omega\rangle_T:=
\sum_{xy\in \cl}\rho(xy)\,\lambda^T_{xy}(\omega).
$
\epf

We will say that an edge-weight (or a strong) lasso $\cl$ for $T$ is {\em tight} if the number of cords in $\cl$
coincides with
the number $|E|$ of edges of $T$ or, equivalently, if the bilinear map `$\langle ... |...\rangle_T $' defines a proper non-degenerate pairing between $\rr^\cl$ and  $\rr^E$, i.e.,    it identifies each of these two vector spaces with the dual of the other.\\

We now show that $\cl$ is connected if $\cl$ is a topological lasso for $T$, and that
it is `strongly non-bipartite' --- i.e.,   every connected
component of the graph $\Gamma(\cl)$ is not bipartite --- if $\cl$ is
an edge-weight lasso for $T$:

\begin{theorem}\label{connectedthm1}
\mbox{ }

{\rm (i)} If $n\ge 4$ holds and $\cl$ is a topological lasso for $T$, then $\cl$ must be connected.

{\rm (ii)} If $\cl$ is  an edge-weight lasso for $T$, then $\cl$ must be strongly non-bipartite.

{\rm (iii)} In particular, $\cl$ must be connected and non-bipartite if $\cl$ is a strong lasso for $T$.

\end{theorem}

\pf (i) Suppose there exists a bipartition of $X$ into two non-empty disjoint subsets $A$ and $B$ such that $\cl$ contains no cord of the form
$ ab$ with $a\in A$ and $b\in B$. Consider any proper edge weighting $\omega$ of $T$, the two trees $T|_A$ and  $T|_B$ obtained by restricting $T$ to $A$ and $B$, respectively, and the associated edge weightings $\omega|_A$ and  $\omega|_B$ of  $T|_A$ and  $T|_B$. Obviously, we can always form
an $X-$tree $T'$ with a proper edge weighting $\omega'$
such that $T'$ is not equivalent to $T$ while
 $T|_A=T'|_A$ and  $T|_B=T'|_B$ as well as
 $\omega|_A=\omega'|_A$ and  $\omega|_B=\omega'|_B$ and, therefore, also
 $(T,\omega)\eb (T',\omega')$  holds, for example by
 `fusing'  $T|_A$ and $T|_B$ via
 any appropriately
chosen bridge.

 (ii)  Suppose that $\cl$ contains a connected component that is bipartite relative to some bipartition of its vertex set $Y$ into the two subsets
 $Y^+$ and $Y^-$.
 Then,
 the set $\cl$ can never be an edge-weight lasso for $T$ as, given any proper edge weighting of $T$ with positive weights on all pendant edges, one can always
add some small constant $\tau$ to the weights of all pendant edges containing a leaf from $Y^+$ and subtract the same amount from the weights of all pendant edges containing a leaf from $Y^-$ without changing the distances between any two leaves $x,x' \in X$ with $xx'\in \cl$.

(iii) The last assertion is a trivial consequence of the first two assertions.
\epf

{\bf Definition:}
Given any cord $\cy=xx'\in \cl$, let $\Gamma(\cl,\cy)=(X-\cy,\cl^{(\cy)})$ denote the sub-graph of $\Gamma(\cl)=(X, \cl)$ with vertex set $X-\cy$ and edge set
$$
\cl^{(\cy)}:=\{yy'\in \cl:yy'\subseteq X- \cy, xx'\|yy'\in \Q(T), \text{ and }xy,x'y'\in \cl \text{ or }xy',x'y\in \cl \}.
$$
\hfill$\Box$

Clearly, given an  edge-weighted  $X-$tree $(T',\omega')$ with $(T,\omega)\eb (T',\omega')$, one has
\begin{equation}\label{'}
 D_\omega(xx'|yy')=D_{\omega'}(xx'|yy')
\end{equation} for any two distinct elements $y,y'\in X- \cy$ with $yy'\in \cl^{(\cy)}$ as if, say, $xy,x'y'\in \cl$ holds, one has
$D_{\omega'}(x,y)+D_{\omega'}(x',y')=D_\omega(x,y)+D_\omega(x',y')>D_\omega(x,x')+D_\omega(y,y')
=D_{\omega'}(x,x')+D_{\omega'}(y,y')$. We claim:

\begin{theorem}\label{top}
Consider a subset $\cl$ of $\binom{X}{2}$ with $X=\bigcup\cl$ and a cord 
$\cy=xx'\in \cl$, and assume that 
$T$ is an $X-$tree. Assume further that the restriction $\Gamma(\cl,\cy)|_A$ of $\Gamma(\cl,\cy)$ to any subset $A$ of $X-\cy$ for which $A,X-A$ is a virtual $T-$split is connected, that $\omega$ is a proper edge weighting of $T$, and that $T'$ is another $X-$tree  with a proper edge weighting $\omega'$ such that $(T,\omega)\eb(T',\omega')$ holds. 
Then, 
$D_\omega(xx'|yy')=D_{\omega'}(xx'|yy')$ must hold for any two distinct elements $y,y'\in X-\cy$ with
$xx'\|yy'\in \Q(T)$.

In particular, the subset $\cl$ of $\ch$ must be a topological lasso for $T$ if the two elements $x,x'$ in $\cy$ form a proper 
$T-$cherry
and $\Gamma(\cl,\cy)|_A$ is connected for any subset $A$ of $X-\cy$ for which $A,X-A$ is a virtual $T-$split.
\end{theorem}
\pf 
 Consider two distinct elements $y,y'\in X- \cy$ with $xx'\|yy'\in \Q(T)$. To show that $D_\omega(xx'|yy')=D_{\omega'}(xx'|yy')$ holds, we will use induction relative to the cardinality of the union $A$ of the two disjoint and non-empty subsets $A(y|y'x)$ and $A(y'|yx)$
for which, according to {\bf 2.3.7}, the split $A,X-A$ is a virtual $T$-split implying that, in view of our assumptions, the restriction $\Gamma(\cl,\cy)|_A$ of $\Gamma(\cl,\cy)$ to $A\subset X-\cy$ is connected.
If $|A|=2$ holds, 
this implies that $yy'\in \cl^{(\cy)}$ and, therefore, also $D_\omega(xx'|yy')=D_{\omega'}(xx'|yy')$ must hold.

Otherwise, our induction hypothesis implies that  $D_\omega(xx'|ya)=D_{\omega'}(xx'|ya)$ holds for all $a\in X-\{x,x',y,y'\}$ with $ya\|xy'\in \Q(T)$, and 
that
 $D_\omega(xx'|y'a')=D_{\omega'}(xx'|y'a')$ 
 holds
 for all $a'\in X-\{x,x',y,y'\}$ with $y'a'\|xy\in \Q(T)$. Furthermore, our assumption that $\Gamma(\cl,\cy)|_A$ is connected now implies that there must exist some cord $aa'\in \cl^{(\cy)}$ and, therefore, $D_\omega(xx'|aa')=D_{\omega'}(xx'|aa')$ with $a\in A(y|y'x)$ and $a'\in A(y'|yx)$. So, our claim holds in case $a=y$ and $a'=y'$. If, say, $a\neq y$ and $a'=y'$ holds, our induction hypothesis implies
$D_{\omega'}(xx'|ya)=D_\omega(xx'|ya)>D_\omega(xx'|ay') =D_{\omega'}(xx'|ay')$ and, therefore,
$D_\omega(xx'|yy')=D_\omega(xx'|ay')=D_{\omega'}(xx'|ay')=D_{\omega'}(xx'|yy')$ in view of (\ref{=}). And if $a\neq y$ and $a'\neq y'$ holds, our induction hypothesis implies
$D_{\omega'}(xx'|ya)=D_\omega(xx'|ya)>D_\omega(xx'|aa') =D_{\omega'}(xx'|aa')$ and $D_{\omega'}(xx'|y'a')=D_\omega(xx'|y'a')>D_\omega(xx'|aa') =D_{\omega'}(xx'|aa')$ and, therefore,
$D_\omega(xx'|yy')=D_\omega(xx'|aa')=D_{\omega'}(xx'|aa')=D_{\omega'}(xx'|yy')$ in view of (\ref{==}), as claimed.

In particular, if the two elements $x,x'$ in $\cy$ form a proper 
$T-$cherry,
they must also form one in $T'$, and a bipartition $A,B$ of $X$ with, say, $x\in B$ forms a non-trivial $T-$split if and only if $B$ contains also $x'$ and 
$D_\omega(xx'|aa')>D_\omega(xx'|ab)$ or, equivalently, $D_{\omega'}(xx'|aa')>D_{\omega'}(xx'|ab)$ 
holds for all $a, a'\in A$ and $b\in B$, that is, if and only if $A,B$ forms a non-trivial $T'-$split. So, $T\simeq T'$
must clearly hold in this case in view of 
the last remark in
{\bf 2.3.7}.  Remarkably, requiring only that $\Gamma(\cl,\cy)|_A$ is connected in case $A,X-A$ is a $T-$split, does not even imply that $\cl$ is a weak lasso for $T$.
\epf

To conclude this section, we now discuss two  instructive examples:
We have seen above that there exist topological lassos for $X-$trees 
(e.g., the tree $T_4$) that 
are not edge-weight lassos.
To show that, conversely, there exist edge-weight lassos for $X-$trees that are not topological lassos, consider the 
`star tree'
$$
T^*=\left(V^*:=X\cup\{*\},E^*:=\big\{\{*,x\}:x\in X\big\} \right)
$$
with leaf set $X$ and exactly one `central' vertex `$*$' of degree $n\ge 3$ adjacent to all leaves of $T^*$.
While it is obvious that any subset of $\ch$, even the empty set, is a topological lasso for the star tree $T^*$ in case $n=3$, there is only one topological lasso in case $n\ge 4$, viz., the set $\binom{X}{2}$:
Indeed, if $\omega$ is, e.g., the `all-one' map ${\bf 1}^{E^*}$
and if
some cord $ab \in \binom{X}{2}$ is not contained in a subset $\cl$ of $\ch$, we may ``extract'' the two leaves in that cord to form a proper cherry that is attached to a vertex $v$ of degree $3$ that in turn is attached to the central vertex $*$ of $T^*$ and adjust the edge length accordingly by putting, say, $\omega'(\{a,v\})=\omega'(\{b,v\})=
\omega'(\{v,*\})=0.5$ and $\omega'(\{x,*\})=1$
for all $x\in X-\{a,b\}$
to obtain an $X-$tree $T'$ with an edge weighting
$\omega'$ for which
$(T^*,\omega)\eb (T',\omega')$ holds. So, no proper subset of
$\binom{X}{2}$ can be a topological lasso -- and, hence, even less a strong lasso -- for $T^*$.

In contrast, it is easy to see that a subset $\cl$ of $\ch$ with $X=\bigcup\cl$ is an edge-weight lasso for $T^*$ if and only if it is strongly non-bipartite implying that -- in accordance with Theorem \ref{LA} -- any edge-weight lasso for $T^*$ contains at least $n$ cords and that all minimal edge-weight lassos for $T^*$ are tight, i.e.,  they contain exactly $n$ cords: Indeed, it follows from Theorem \ref{connectedthm1} (ii) that any edge-weight lasso for any $X-$tree must be strongly non-bipartite. And, conversely, if a subset $\cl$ of $\ch$ is strongly non-bipartite, there exists, for every $x\in X$, some sequence
$x_0:=x,x_1,x_2 \dots,x_{2k},x_{2k+1}:=x$, $k \geq 0$, consisting of elements from $X$ such that
$x_ix_{i+1}\in \cl$ holds for all $i=0,\dots, 2k$. Thus, if
$\omega,\omega'$ are any two edge weightings of $T^*$ with $(T^*,\omega)\eb (T^*,\omega')$, the sum
$\sum_{i=0}^{2k}(-1)^i D_{\omega}(x_i,x_{i+1})=\sum_{i=0}^{2k}(-1)^i (\omega(e_{x_i})+
\omega(e_{x_{i+1}}))=2\omega(e_x)$ must coincide with the sum $\sum_{i=0}^{2k}(-1)^i D_{\omega'}(x_i,x_{i+1})=2\omega'(e_x)$ implying that $\omega$ and $\omega'$ must coincide on all edges of $T^*$.

\smallskip
It follows in particular that, in contrast to Assertion (i) in Theorem \ref{connectedthm1}, $\cl$ can be disconnected if $\cl$ is merely an edge-weight lasso for a (non-binary) $X-$tree. An example is provided by
the star tree with the leaf set 
$X_6:= \{a,b,c,d,e,f\}$ 
and the set $\cl := \binom{\{a,b,c\}}{2} \cup \binom{\{d,e,f\}}{2}$ which, consisting of two disjoint triangles,
is clearly strongly non-bipartite.
However, we will see shortly (Corollary~\ref{connectedcor})
that $\cl$ must be connected whenever $\cl$ is an edge-weight lasso for a binary $X-$tree.

\begin{figure}[ht]
\begin{center}
\resizebox{6cm}{!}{
\includegraphics{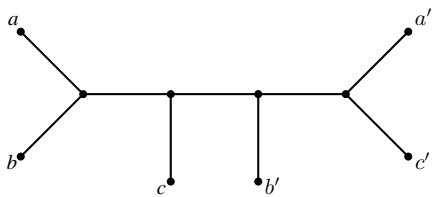}
}
\caption{The binary $\{a,b,c,a',b',c'\}-$tree $T_6$
for which $\cl_6$ 
as defined in (\ref{cl6eq})
 is an edge-weight, but not a topological lasso.}
\label{figure_caterpillar}
\end{center}
\end{figure}

\medskip
Finally, we show that there exist edge-weight lassos also for binary $X-$trees that are not topological lassos:
Consider the set $X_6':=\{a,b,c,a',b',c'\}$, the binary 
$X_6'-$tree 
$T_6$ depicted in Fig. \ref{figure_caterpillar},
and the subset
\begin{equation}
\label{cl6eq}
\cl_6:= \binom{ \{ a,b,c\}}{2}\cup\binom{ \{ a',b',c'\}}{2}\cup \big\{ aa',bb',cc'\big\}
\end{equation}
of $\binom{X_6'}{2}$.
 $\cl_6$
is an edge-weight lasso for $T_6$, since we can 
determine, for any proper edge weighting  $\omega$ of $T_6$, 
 the values of $D(x,y)$ for the metric
$D:=D_\omega$ 
for the six `missing' cords $xy$ in $\binom{X_6'}{2}- \cl_6$ starting from the $D$--values of
the cords in $\cl_6$. For example,
we have
$D(a,b')=D(a,c)+D(b,b')-D(b,c)$, from which we can
compute $D(b,a')$ as
$D(b,a')=D(a,a')+D(b,b')-D(a,b')$.
By symmetry, we can also compute $D(c,a')$ directly from the data and, then, $D(a,c')$ which, finally, allows us to also compute $D(b,c')=D(b,a')+D(a,c')-D(a,a')$ and $D(c,b')=D(c,a')+D(a,b')-D(a,a')$.

\begin{figure}[ht]
\begin{center}
\resizebox{8cm}{!}{
\includegraphics{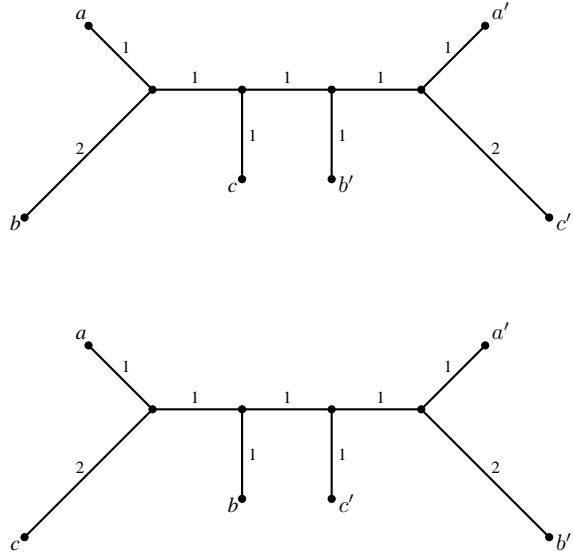}
}
\caption{Although $\cl_6$ is an edge-weight lasso for $T_6$, it fails to be a strong lasso since 
both of the two edge-weighted trees depicted above induce the same distances on all cords in $\cl_6$.
}
\label{figure_not_lasso}
\end{center}
\end{figure}

However, the example in Fig.~\ref{figure_not_lasso} shows that $\cl_6$ does not lasso the shape of $T_6$ 
and, so, is 
not a strong lasso for $T_6$.

\section{Towards a recursive analysis of lasso sets}
\label{rec}

In this section, we will establish a result that can be used to analyse lassos  recursively: Given an $X-$tree $T=(V,E)$, we define a non-empty subset $U$ of $V$ to be a {\em$T$-core}  if the induced subgraph
$
T_U:=(U,E_U:=\{e \in E: e \subseteq U\})
$
of $T$ with vertex set $U$
is connected (and, hence, a tree) and the degree
$\deg_{T_U}(v)$
of any vertex $v$ in $T_U$ is either $1$ or coincides with the degree
$\deg_T(v)$ of $v$ in $T$.
Clearly, putting $E_v=E_v^T:=\{e\in E: v \in e\}$ and
$N_v=N_v^T:=\cup_{e\in E_v}e$ for every vertex $v$ of $T$, $N_v$ is a $T$-core for every $v\in V$, and so is
$\bigcup_{v\in U}N_v$ for every  subset $U$ of $V$ for which $T_U$ is connected.

It is also obvious that
$T_U$ must be an $X_U-$tree for
$X_U:=\{y\in U: \deg_{T_U}(y) =1\}$
for every $T$-core $U\subseteq V$ as $X_U$ is the leaf set of $T_U$.

Further, let $x_U$ denote, for any leaf $x\in X$ of $T$ and any
$T$-core $U\subseteq V$, the
{\em gate} of $x$ in $U$, i.e.,  the unique vertex in $U$ that is closest
to $x$.
Note that $X_U$ must coincide with the set
$gate_U(X):=\{x_U: x\in X\}$ of all gates of the elements of $X$ in $U$.

Finally, for any subset $\cl$ of $\binom{X}{2}$, let $\cl_U$ denote the set consisting of all pairs of distinct elements $y ,y'$ in $X_U$ for which there exists a cord $xx'\in \cl$ with
$x_U=y$ and $x'_U=y'$.  Then, the following holds:

\begin{theorem}\label{core}
Given an $X-$tree $T=(V,E)$, a $T$-core $U\subseteq V$, and a subset $\cl$ that
is a weak, an edge-weight, a topological, or a strong lasso for $T$. Then, the set
 $\cl_U$ is, respectively, a weak, an edge-weight, a topological, or a strong lasso for $T_U$.
 In particular, the graph $\Gamma(\cl_U)=(X_U,\cl_U)$ must be strongly
non-bipartite for every $T$-core $U\subseteq V$ whenever $\cl$ is an
edge-weight lasso for $T$.
  \end{theorem}
\pf
Indeed, this is a direct consequence of the following simple observation:
Given any $X_U-$tree $T'=(U',F')$ with $U'\cap V= X_U$, the graph $T^*$ with vertex set $V^*:=(V-U)\cup U'$ and edge set $E^*:=(E-E_U)\cup F'$ obtained by replacing the interior vertices and edges of $T_U$ in $T$ by those of $T'$ is a well-defined $X-$tree. Furthermore, one has
$$E_{T_U}(x_U|y_U)\subseteq E_T(x|y), \mbox{ }
E_{T'}(x_U|y_U)\subseteq E_{T^*}(x|y),$$ and
$$E_T(x|y)-E_{T_U}(x_U|y_U)=E_{T'}(x_U|y_U)- E_{T^*}(x|y),$$ for all $x,y\in X$.
So, restricting any edge weighting $\omega$ of $T$ to the edges of $T_U$ induces an edge weighting $\omega_U$ of $T_U$,
and extending any edge weighting $\omega'$ of $T'$ to an  edge weighting $\omega^*$ of $T^*$ by putting $\omega^*(e):=\omega(e)$ for every $e\in E\cap E^*=E-E_U$ yields pairs of edge-weighted $X-$ and $X_U-$trees such that
$(T^*, \omega^*)\eb (T, \omega)$ if and only if $(T', \omega') \eu (T_U, \omega_U)$.
\epf

This theorem has a simple, yet useful consequence. To describe it, we define the following graph.

{\bf Definition:}
Given a non-empty subset $\cl$ of $\binom{X}{2}$, an $X-$tree $T$, and a vertex $v$ of $T$, let
$G(\cl,v)$ denote the graph with vertex set $E_v$ and edge set $\E_{\cl,v}$ consisting of all pairs $\{e,e'\}\in\binom{E_v}{2}$ for which some  cord $xy$ in $\cl$ with $e,e'\in E_T(x|y)$ exists.
\hfill$\Box$

Given any interior vertex $v\in V$, we can apply
Theorem \ref{core} to the $T$-core $U:=N_v$ which, together with our results on star trees, yields:
\begin{corollary}\label{one-point}
Given any interior vertex $v\in V$ and any edge-weight lasso $\cl$ for $T$, the graph $G(\cl,v)$  is strongly non-bipartite and is therefore the complete graph with vertex set $E_v$ if
$\deg_T(v)=3$.
It is also the complete graph with vertex set $E_v$ independently of the degree of $v$ when $\cl$ is a topological lasso for $T$.
\end{corollary}
It follows that a necessary condition for a subset $\cl$ of $\ch$ to lasso the edge weights or the shape of an $X-$tree $T$ is that the graph $G(\cl,v)$ is strongly non-bipartite or
the complete graph with vertex set $E_v$, respectively, for every interior vertex $v$ of $T$.

\subsection{The case of a proper cherry $a,b$.}
\label{cherrysec}

Let us now suppose  that $a,b$ is a proper 
 $T-$cherry for some
$X-$tree $T = (V ,E)$ and let
$e_{ab} \in E$ denote the unique interior edge of $T$ that is adjacent
to $v:=v_a=v_b$ so that
$E_v = \{e_a, e_b , e_{ab}\}$ holds.
Note that the set $U=U_{ab}:=V-\{a,b\}$ obtained by deleting the two leaves $a,b$ in the vertex set $V$ of $T$ is a $T$-core, and that the leaf set $X_U$ of the associated tree $T_U$ with vertex set $U$ coincides with the set $(X-\{a,b\})\cup\{v\}$. 
Note also
that $e_{ab}$ is the unique pendant edge $e_v^{T_U}$ of $T_U$ containing its leaf $v$, and that
$v = a_U = b_U, E_U = E - \{e_a , e_b \}$, and
$|E_U | = |E| - 2$ holds.\\

We pause to introduce some further terminology.  Given any non-empty subset $\cl$ of $\ch$,
put $\cl^{ab}:=\cl-\{ab\}$ and, given in addition any two distinct elements $x,y\in X$,
put
$$
\delta_{xy\in \cl}:=\begin{cases}
      & 1, \text{ if } xy\in \cl; \\
      & 0, \text{ otherwise.}
\end{cases}
$$
Further,  let $X(x,\cl)$ denote the set
$$X(x,\cl):=\{y\in X: xy\in \cl\}.$$

If $\omega_U$ denotes the restriction of an edge weighting $\omega \in \Omega_T$ to $E_U$, the linear maps
$\lambda_{xy}^T$ introduced above (Eqn. \ref{lambdaeq}) satisfy:
\begin{equation}\label{abx}
\lambda_{ax}^{T}(\omega)=\omega(e_a) +\lambda_{vx}^{T_U}(\omega_U)
\text{ \,\,and\,\, }
\lambda_{bx}^{T}(\omega)=\omega(e_b) +\lambda_{vx}^{T_U}(\omega_U)
 \end{equation}
for all $x\in X- \{a,b\}$, and
$
\lambda_{xy}^{T}(\omega)=\lambda_{xy}^{T_U}(\omega_U)
$
for every cord $xy\in \binom{X- \{a,b\}}{2}$.
We also have
$$
\cl_U= (\cl\cap \binom{ X- \{a,b\} }{2})
\cup \big\{\{x,v\}:x\in X(a,\cl^{ab})\cup X(b,\cl^{ab})\big\}
$$
and, therefore, also
 \begin{equation}\label{|cbU|}
 |\cl_U|=|\cl|-\delta_{ab\in \cl}-|X(a,\cl^{ab})\cap X(b,\cl^{ab})|
\end{equation}
 for every subset $\cl$ of $\ch$.
Thus, if $\cl$ is a tight edge-weight lasso for $T$, we must have
 \begin{eqnarray*} &&|\cl|=|E|=|E_U|+2\\ &&\le |\cl_U|+2=|\cl|-\delta_{ab\in \cl}-|X(a,\cl^{ab})\cap X(b, \cl^{ab})|+2\\ &&= |\cl|+1-|X(a,\cl^{ab})\cap X(b,\cl^{ab})|\le |\cl|+1
  \end{eqnarray*}
  because  (i) $\cl_U$ must be an edge-weight lasso for $T_U$ according to Theorem \ref{core} and (ii) $ab\in \cl$ must hold in this case.

In consequence, the induced edge-weight lasso $\cl_U$ for $T_U$ is tight if and only if there exists some (necessarily unique!) leaf $x\in X-\{a,b\}$ with $xa,xb\in \cl$.
Otherwise,  $X(a,\cl^{ab})\cap X(b,\cl^{ab})=\0$ must hold and $\cl_U$ has cardinality $|E_U|+1$ in which case there must exist -- up to scaling -- exactly one non-zero map $\rho_U: {X_U\choose 2}\ra \rr$ with support in $\cl_U$ and
 $$\sum_{xy\in {X_U\choose 2}}\rho_U(x,y)
 \lambda^{T_U}_{xy}=0.
 $$

Furthermore, we must have $\alpha:=\sum_{x\in X(a,\cl^{ab})}\rho_U(x,v)\neq 0$ and $\beta:=\sum_{x\in X(b,\cl^{ab})}\rho_U(x,v)\neq 0$ for every such non-zero map $\rho_U$:
Indeed, put
$$\rho(x,y):=
\begin{cases}
      & \rho_U\big(x_U, y_U\big), \text{ if } xy \in \cl; \\
      & 0, \text{ if }  xy\in \ch-\cl;
\end{cases}
$$
and note that $\rho$ is a non-zero map with support in $\cl$. Note also that -- as $\cl_U$ is, by assumption, the disjoint union of $\big\{\{v,x\}:x\in X(a,\cl^{ab})\big\}$, $\big\{\{v,x\}:x\in X(b, \cl^{ab})\big\}$, and $\cl_U\cap {X-\{a,b\}\choose 2}$ -- also 
\begin{eqnarray*} 
&&\sum_{xy\in \ch}\rho(x,y)\,\lambda^{T}_{xy}(\omega) \\ 
&=&  \sum_{xy\in {X-\{a,b\}\choose 2}}\rho_U(x,y)\, \lambda^{T_U}_{xy}(\omega_U) +\sum_{x\in X(a,\cl^{ab})}\rho(a,x)\big(\lambda^{T_U}_{xv}(\omega_U)+\omega(e_a) \big)\\ 
&&+ \sum_{x\in X(b,\cl^{ab})}\rho(b,x)\big(\lambda^{T_U}_{xv}(\omega_U)+\omega(e_b) \big)\\ 
&=&  \sum_{xy\in {X_U\choose 2}}\rho_U(x,y)\,\lambda^{T_U}_{xy} (\omega_U) +\alpha\,\, \omega(e_a)+\beta\,\,\omega(e_b)\\ 
\end{eqnarray*} 
must hold for every map $\omega\in \rr^E$. However, noting that $\sum_{xy\in {X_U\choose 2}}\rho_U(x,y)\, \lambda^{T_U}_{xy}$ vanishes by our choice of $\rho_U$, we must \big(with $\delta_{e_{ab}}$ as defined by Eqn. (\ref {eeq}) \big) also have
\begin{eqnarray*} 0&=&\sum_{xy\in {X_U\choose 2}}\rho_U(x,y)\,\lambda^{T_U}_{xy} (\delta_{e_{ab}})\\ &=& \sum_{x\in X(a,\cl^{ab})}\rho_U(x,v)+\sum_{x\in X(b,\cl^{ab})}\rho_U(x,v)\\ &=&\alpha+\beta \end{eqnarray*} 
So, $\alpha=0$ would imply that also $\beta=0$ must hold and, therefore, also 
$$ 
\sum_{xy\in \ch}\rho(x,y)\,\lambda^{T}_{xy}(\omega)
=\sum_{xy\in {X_U\choose 2}}\rho_U(x,y)\,\lambda^{T_U}_{xy} (\omega_U)
 =0
$$ 
for every map $\omega\in \rr^E$, which is impossible if $\cl$ is a tight lasso for $T$.

This yields a good part of the following result.

\begin{theorem}\label{cbU}
Continuing with the definitions and notations introduced at the start of Section {\rm \ref{cherrysec}}, a subset $\cl$ of $\ch$  is an edge-weight lasso for $T$ if and only if:

\noindent
$({\bf  U1})$ $\cl$ contains the cord $ab$;

\noindent
$({\bf  U2})$ $\cl_U$ is an edge-weight lasso for $T_U$;

\noindent
and  at least one of the following two assertions $({\bf  U3-a})$ or $({\bf  U3-b})$  holds:

 \noindent
$({\bf  U3-a})$ The two subsets $X(a,\cl^{ab})$ and $X(b,\cl^{ab})$ of $X$ have a non-empty intersection.

 \noindent
 $({\bf  U3-b})$ There exists some
  non-zero map $\rho_U: {X_U\choose 2}\ra \rr$  with support in $\cl_U$ and 
$$
\sum_{xy\in \cl_U}
\rho_U(x,y)
\lambda^{T_U}_{xy}=0 \text{ as well as }\sum_{x\in X(a,\cl^{ab})}\rho_U(x,u)\neq 0. 
$$

In particular, a subset $\cl$ of $\ch$ is a tight edge-weight lasso for $T$ if and only if it has cardinality $|E|$, and $({\bf  U1})$, $({\bf  U2})$,
and either one of the following two assertions $({\bf  U3-a'})$ or $({\bf  U3-b'})$ holds:

 \noindent
$({\bf  U3-a'})$ $\cl_U$ is a tight edge-weight lasso for $T_U$,

 \noindent
 $({\bf  U3-b'})$ $\cl_U$ has cardinality $|E_U|+1=|\cl|-1$ and $\sum_{x\in
X(a,\cl^{ab})}\rho_U(x,u)\neq 0$ holds for every non-zero map $\rho_U: {X_U\choose 2}\ra \rr$ with support in $\cl_U$ for which
 $\sum_{xy\in \cl_U}
 \rho_U(x,y)
 \lambda^{T_U}_{xy}=0$ holds.
\end{theorem}

\pf In view of our observations above applied to any tight lasso for $T=(V,E)$ contained in $\cl$, it suffices to show that a subset $\cl$ of $\ch$ is an edge-weight lasso for $T$ if $({\bf  U1})$,  $({\bf  U2})$ and at least one of the assertions $({\bf  U3-a})$ or $({\bf  U3-b})$ hold.

So, assume that, for some map $\eta\in \rr^E$, one has $\lambda_{xy}^{T}(\eta)=0$ for all cords $xy\in \cl$. We have to show that $\eta(e)=0$ must hold for every edge $e\in E$.
To this end, note first that  to establish our claim, it suffices, in view of (\ref{abx}), to show that, if ({\bf  U1}), ({\bf  U2}), and either ({\bf  U3-a}) or  ({\bf  U3-b}) hold, then $\eta(e_a)=\eta(e_b)=0$ must hold for every map
$\eta\in \rr^E$ as above.

Yet, if ({\bf U3-a}) holds (i.e., if
$xa,xb\in \cl$ holds for some $x \in X- \{a,b\}$), the assumption that
 $\lambda_{xy}^{T}(\eta)=0$ holds for some $\eta\in \rr^E$ and for all cords $xy\in \cl$ implies that the following hold:

 \begin{eqnarray*}
0&=&\lambda_{ab}^{T}(\eta)=\eta(e_a) +\eta(e_b),\\
0&=&\lambda_{ax}^{T}(\eta)=\eta(e_a) +\lambda_{ux}^{T}(\eta_U),\\
0&=&\lambda_{bx}^{T}(\eta)=\eta(e_b) +\lambda_{ux}^{T}(\eta_U).
\end{eqnarray*}
This readily implies $\eta(e_a)=\eta(e_b)=0$ in this case (since we may add the first equation to either the second or the third one, and subtract the other one).

Moreover,  if ({\bf  U1}) and ({\bf  U2}) hold,
if $\sum_{xy\in {X_U\choose 2}}\rho_U(x,y)\lambda^{T_U}_{xy}=0$
and also $\sum_{x\in  X(a,\cl^{ab})}\rho_U(x,u)\neq 0$ holds for some non-zero map
$\rho_U: {X_U\choose 2}\ra \rr$,
and if $\lambda_{xy}^{T}(\eta)=0$ holds for all $xy\in \cl$ for some map $\eta\in \rr^E$, we must have
$$
\lambda_{ux}^{T_U}(\eta_U)=-\eta(e_a) \text{ for all }x \in X(a,\cl^{ab}),
$$
 $$
 \lambda_{ux}^{T_U}(\eta_U)=-\eta(e_b)
 \text{ for all } x \in X(b,\cl^{ab}),
 $$
 and
 $$
 \rho_U(x,y)\lambda_{xy}^{T_U}(\eta_U)=0
  \text{ for all } xy\in {X-\{a,b\}\choose 2}.
  $$
   Thus, evaluating the identity  $\sum_{xy\in \cl_U}\rho_U(x,y)\lambda^{T_U}_{xy}=0$ on $\eta_U$ and noting that $\cl_U$ is, by assumption, the disjoint union of $\big\{\{v,x\}:x\in X(a,\cl^{ab})\big\}$, $\big\{\{v,x\}:x\in  X(b,\cl^{ab})\big\}$, and $\cl_U\cap {X-\{a,b\}\choose 2}$, we get:
\begin{eqnarray*}
0&=&\sum_{x   \in X(a,\cl^{ab})}-
\rho_U(v,x)
\eta(e_a)
+\sum_{x   \in X(b,\cl^{ab})}-
\rho_U(v,x)
\eta(e_b) \\
&=&-\eta(e_a) \big(\sum_{x   \in X(a,\cl^{ab})}
\rho_U(v,x)
 \big)
   -\eta(e_b)\big(\sum_{x   \in X(b,\cl^{ab})}
\rho_U(v,x)
\big).
\end{eqnarray*}
Furthermore, evaluating the identity $\sum_{xy\in {X_U\choose 2}}\rho_U(x,y)\lambda^{T_U}_{xy}=0$ on the map $\delta_{e_{ab}}$ (Eqn. \ref{eeq})  also yields the following:
$$
0=\sum_{x   \in X-\{a,b\}}\rho_U(v,x)=
\sum_{x   \in X(a,\cl^{ab})}\rho_U(v,x)+
\sum_{x   \in X(b,\cl^{ab})}\rho_U(v,x)
$$
and, therefore, the following holds:
$$
0=\big(\eta(e_b)-\eta(e_a) \big)\sum_{x   \in X(a,\cl^{ab})}\rho_U(u,x).
$$
So, our assumption $\sum_{x\in
X(a,\cl^{ab})}\rho_U(x,v)\neq 0$ implies
$0=\eta(e_b)-\eta(e_a)$ which, together with $0=\lambda_{ab}^{T}(\eta)=\eta(e_a) +\eta(e_b)$, implies that also in this case  $0=\eta(e_a) =\eta(e_b)$ must hold, as claimed.\epf

Our observations imply also that we can construct all the tight edge-weight lassos of $T$ from the
 edge-weight lassos $\cl'$ of $T_U$ with
 $|\cl'|\le |E_U|+1$ as follows:

\begin{corollary}\label{construction}
\mbox{}
$(i)$ Given any tight edge-weight lasso $\cl'$ of $T_U$, there is a canonical one-to-one correspondence between all tight edge-weight lassos $\cl$ of $T$ with $\cl_U=\cl'$ and all pairs of subsets $A,B$ of
$X_U(v,\cl')$ with $|A\cap B|=1$.

$(ii)$ Furthermore, given any edge-weight lasso $\cl'$ of $T_U$ of cardinality $|E_U|+1$, there is a canonical one-to-one correspondence between all tight edge-weight  lassos $\cl$ of $T$ with $\cl_U=\cl'$ and all pairs of disjoint subsets $A,B$ of
$X_U(v,\cl')$ for which $\sum_{x\in
A}\rho_U(x,v)\neq 0$ holds for one $($or, equivalently, every$)$  non-zero map
$\rho_U: {X_U\choose 2}\ra \rr$ with support in $\cl'$ for which
 $\sum_{xy\in {X_U\choose 2}}\rho_U(x,y)\lambda^{T_U}_{xy}=0$ holds.

 In both cases, the correspondence is given by associating to each pair $A,B$, the set:
 $$
 \cl'_{A,B}:=(\cl'\cap {X-\{a,b\}\choose 2})
 \cup \{ax: x\in A\cup\{b\}\}\cup \{bx: x\in B\}.
 $$
\end{corollary}

As a second consequence of Theorem~\ref{cbU}, we have the following result already indicated in the remark at the end of Section \ref{facts}.
\begin{corollary}
\label{connectedcor}
If $\cl$ is an edge-weight lasso for a binary $X-$tree, then $(X, \cl)$ is a connected graph.
\end{corollary}

\section{Shellability}
\label{shell}

In this section, we introduce a concept that relates to strong lassos and
will apply in particular in the discussion of all edge-weight lassos
for $X-$trees with $|X|=5$ and other examples
in \cite{dres2}:   Given a subset $\cl$ of $\ch$ with $X=\bigcup \cl$, and an $X-$tree $T$, we say that $\ch-\cl$ is {\em $T$--shellable} if
there exists a labelling of the cords in $\ch-\cl$ as, say, $a_1b_1, a_2b_2, \dots, a_mb_m$ such that, for every $\mu\in \{1,2,\dots,m\}$, there exists a pair $x_\mu,y_\mu$ of
 `pivots' for $a_\mu b_\mu$, i.e.,  two distinct elements $x_\mu,y_\mu\in X-\{a_\mu,b_\mu\}$,
 for which the tree $T|_{Y_\mu}$ obtained from $T$ by restriction
to $Y_\mu:=\{a_\mu,b_\mu,x_\mu,y_\mu\}$, is a quartet tree of type
$a_\mu x_\mu\|y_\mu b_\mu$, and
all cords in $\binom{Y_\mu}{2}$ except $a_\mu b_\mu$ are contained
in $\cl_\mu:=\cl \cup \big\{a_{\mu'}b_{\mu'}: \mu'\in
\{1,2,\dots,\mu-1\}\big\}$.
Any such labelling of $\ch-\cl$ will also be called a $T$-{\em
shelling} of  $\ch-\cl$, and any subset $\cl$ of $\ch$ for which a $T$-
{\em shelling} of  $\ch-\cl$ exists will  also be called an {\em $s-$lasso for  $T$}.

\subsection{Example}
\label{shelex}

Consider the caterpillar tree $T_5$ on $X_5$
depicted in Fig.\ref{fig:induced}.
We claim that
the set $\cl:=\{ab, bc, cd, de, ea, ad, ac\}$, is an $s-$lasso for $T_5$:
Indeed, labelling the elements in the cords in
$\ch-\cl=\{bd,be,ce\}$ as
$$
a_1:=c, b_1:=e;\,\, a_2:=b,b_2:=e; \,\mbox{ and } \, a_3:=b ,b_3:=d
$$
yields a $T_5-$shelling  of $\ch-\cl$ because,
choosing the elements
$$
x_1:=a, y_1:=d;\,\,x_2:=a, y_2:=c; \,\mbox{ and } \, x_3:=a, y_3:=e
$$
as pivots, the quartet trees
$T_5|_{Y_\mu}$ are indeed quartet trees of type
$a_\mu x_\mu\|b_\mu y_\mu$ for
$
Y_\mu=\{a_\mu,b_\mu,x_\mu,y_\mu\},\,\,\,\mu=1,2,3,
$
  as required, and all cords in
$\binom{Y_\mu}{2}$ except $a_\mu b_\mu$ are contained in $\cl_\mu=
\cl \cup \big\{a_{\mu'}b_{\mu'}: \mu'\in  \{1,2,\dots,\mu-1\}\big\}$
for all $\mu=1,2,3$.
Thus, listing the cords in $\ch-\cl$ `anti-lexicographically' in the
order $ce,be,bd$
yields a $T_5-$shelling of ${X\choose 2}-\cl$, implying that $\cl$ is
an $s-$lasso for $T_5$ as claimed.
\hfill$\Box$\\

We now establish the following simple, yet sometimes rather helpful result:
\begin{theorem}\label{shell=>strong}Every $s-$lasso $\cl\subseteq \ch$
for an $X-$tree $T$ is a strong lasso for $T$.
\end{theorem}
 \pf  Given 
 a $T-$shelling $a_1b_1, a_2b_2, \dots, a_mb_m$ of $\ch-\cl$ with corresponding pivots  $x_1,y_1; x_2,y_2;\dots; x_m,y_m$,
 we can compute, for every proper edge weighting $\omega$ of $T$, the distances $D_\omega(a_\mu,b_\mu)$ for all $\mu\in \{1,2,\dots,m\}$ recursively, because:
$$
D_\omega(a_\mu,b_\mu)=D_\omega(a_\mu,y_\mu)+D_\omega(b_\mu,x_\mu)-D_\omega(x_\mu,y_\mu)
$$
must hold in view of the fact that, by assumption,
$$
D_\omega(a_\mu,x_\mu)+D_\omega(b_\mu,y_\mu)< D_\omega(a_\mu,b_\mu)+D_\omega(x_\mu,y_\mu)
$$
must hold.\epf

The converse to Theorem ~\ref{shell=>strong} does not hold, that is,
there exists an $X-$tree $T$ and a strong
lasso for $T$ that is not an $s-$lasso for $T$, as the following example shows.

\subsection{Example}
\label{ex}
Put $X_7: =\{a,b,c,d,e,f,g\}$, and let $T_7$ denote the binary $X_7-$tree 
with exactly two proper cherries 
$a,b$ and $f,g$, and 
the three `single' leaves $c,d,e$. Assume furthermore that 
that the corresponding  adjacent vertices $v_c,v_d,v_e$ are passed
in this order on the path connecting the cherry $a,b$ with $f,g$.   
Then, 
the bipartite set 
$$
\cl_7: = \{ab, ad, bc, be,
cd, cf, de, dg, ef, fg\}
$$ 
is a topological lasso for $T_7$ since
any $X_7-$tree  $T'$ with
  $(T_7, \omega) \ew (T', \omega')$ for some proper edge weightings $
\omega\in \Omega_{T_7}$ and
$\omega'\in \Omega_{T'}$ must display
the quartets
$ab\|cd, bc\|de, cd\|ef,$ and $de\|fg$ which is well known to imply that $T_7$ and
$T'$ must be equivalent (see e.g. \cite{boc} or \cite{Dre10}).   It follows that adding
the cord $ag$ to $\cl_7$ yields an edge-weight lasso for $T_7$, since the associated $11 \times 11$ incidence matrix of paths (one for  each cord) and edges of $T_7$ has full rank.   Thus $\cl=\cl_7\cup\{ag\}$ is a strong lasso for $T_7$ which, however, is easily seen
not to be an $s-$lasso for $T_7$ as there exists not even any
$4$-subset $Y$ of $X_7$ with
$|{Y\choose 2}\cap \cl|\ge 5$.

\section{Covers of binary $X-$trees}
\label{covsec}

\begin{figure}[ht]
\includegraphics[width=1.0\textwidth] {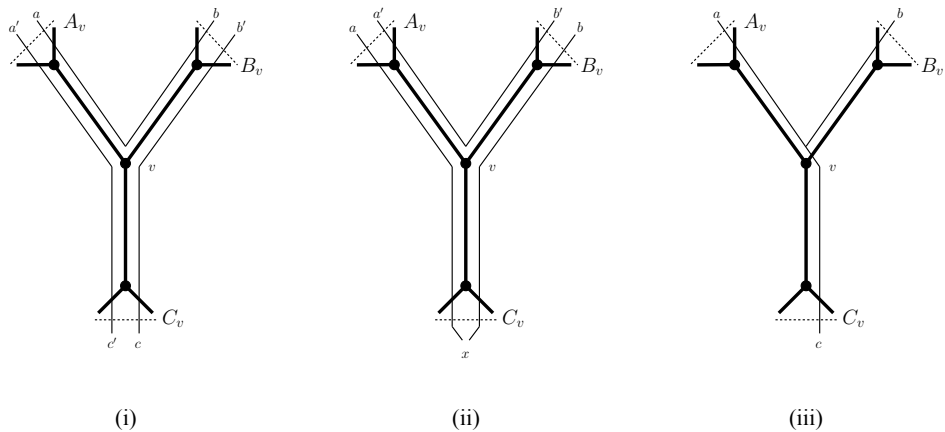}
\caption{(i) The three cords $ab, b'c,$ and $c'a'$ provide a `cover of the
degree  $3$
vertex $v$'; (ii)
the three cords $ax, bx,$ and $a'b'$
provide a `pointed cover of $v$'; (iii)
the three cords $ab, bc,$ and $ca$ provide a `triplet cover of $v$'. It is possible for $a=a'$ or $b=b'$ (in cases (i) and (ii)) and also $c=c'$ (in case (i))
(see text for details).
}
\label{coverfig}
\end{figure}

Recall that, by Corollary~\ref{one-point}, a necessary condition for a subset $\cl$ of $\ch$ to lasso the edge weights
(or, respectively, the shape) of a tree $T$ is that,
for every interior vertex $v$ of $T$,
the graph $G(\cl,v)$ defined above is strongly non-bipartite
(or, respectively, a complete graph).
This suggests the following:

\bigskip
\noindent{\bf Definition:}   A subset $\cl$ of $\binom{X}{2}$ is  an {\em $e-$cover}  of $T$ if
$X$ coincides with $\bigcup \cl$ and $G(\cl,v)$ is strongly non-bipartite for every interior vertex $v$ of $T$, and it is called a {\em $t-$cover}  of $T$ if $X=\bigcup \cl$ holds and $G(\cl,v)$ is a complete graph for every interior vertex $v$ of $T$  -- see Fig. \ref{coverfig} (i) below for an illustration.

\bigskip

In the first part of this section, we restrict our attention to
covers of binary $X-$trees.  Clearly, e- and $t-$covers coincide for such trees  -- so, we will just call them covers in this case.

By definition (and Corollary \ref{one-point}), every edge-weight lasso for an $X-$tree $T$
is an $e-$cover of $T$.
The converse, however, does not hold,  not even for binary $X-$trees.  For example,  the topological lasso
 $\cl_4=\{ab,ac,bd,cd\}$
for the quartet tree $T_4$ of type $ab\|cd$ depicted in Fig.~\ref{figure_quartet} is clearly a cover for $T_4$, but -- in view of $|\cl|<|E|$ -- it does not lasso the edge weights of $T$.

Note 
also that, if $\cl$ is a topological lasso for $T$, then 
$\cl$ must also be a $t-$cover of $T$. However, once again, the converse does not hold; for example,
the 
subset
\begin{equation} 
\label{b3eq}
\cl_6':=\{ab, ac, a'b',a'c', bb', cc'\}
\end{equation}
of $\cl_6$ as defined in (\ref{cl6eq}) 
is a cover for the two non-equivalent binary 
$X_6'-$trees depicted in Fig.~\ref{figure_not_lasso}.

More strikingly, $\cl_6$ itself is, thus, both an edge-weight lasso and a $t-$cover for $T_6$, yet $\cl_6$ fails to lasso the shape of this tree.

We now describe two particular types of covers: Given a binary $X-$tree $T=(V,E)$, we will say that a subset $\cl$ of $\ch$ is a {\em triplet cover} of $T$ if, for every interior vertex $v\in V$ of $T$, there exist three distinct leaves $a,b,c$ with $
ab, ac, bc\in \cl$ and $v={\rm med}_T(a,b,c)$ (see
Fig. \ref{coverfig} (iii) for an illustration of this concept).
Note that a triplet cover of a binary $X-$tree $T$ can be represented as a collection $\cC$ of $3$--element subsets of $X$ with $\bigcup \cC = X$ and with the property that the function
that assigns each triplet to its associated median vertex in  $T$ maps $\mathcal C$ surjectively onto the set of interior vertices of $T$.
A combinatorial characterization of arbitrary collections $\mathcal C$ of $3$--element subsets of $X$ with
$\bigcup {\mathcal C} = X$ for which this function is  injective for some binary $X-$tree is simply, as described recently in
 \cite{dre2}, that $|\bigcup \cC'|
\geq |\cC'|+2$ holds for all non-empty subsets $\cC'$ of $\cC$.

Secondly, given an element $x\in X$, a subset $\cl$ of
${X\choose 2}$ is called a {\em pointed $x-$cover} of a binary phylogenetic $X-$tree $T$ if it is a cover of $T$ and there exist,
for each interior vertex $v$ of $T$, two distinct leaves $a,b\in X$ with $ax, bx\in \cl$ and $v={\rm med}_T(a,b,x)$ (see Fig.~\ref{coverfig} (ii) for an illustration). Moreover,  it is called just a pointed cover of $T$ if there exists some $x\in X$ such that $\cl$ is a pointed $x-$cover of $T$.

 Clearly, every triplet and every pointed cover of $T$ is, in particular, a cover of $T$.
To present some examples of  triplet and pointed covers, recall first that a {\em circular ordering} of (the leaf set $X$ of) an
$X-$tree $T$ is a cyclic permutation $\sigma$ of the elements in $X$ for which there exists a planar embedding of $T$ such that, for every $x\in X$, the leaf that follows the leaf $x$ when one traverses the leaves of $T$ in that embedding in, say, a clockwise fashion is the leaf $\sigma(x)$.
An equivalent characterization is that each edge of $T$ is covered only twice by the
 paths connecting the $n$ pairs of leaves in the set $\big\{\{x, \sigma(x)\}: x\in X\big\}$.
For example,  there exist planar embeddings of the two $X_6'$-trees
depicted in Fig.~\ref{figure_not_lasso} such that
the permutation 
$(a,b,b',a',c',c)$ is a circular ordering for both of them,
while the permutation 
$(a,b,a',c',b',c)$ 
is a circular ordering for $T_6$
(under a different planar embedding), but not for 
the other $X_6'$-tree depicted in that figure
under any planar embedding (as the three paths connecting the three pairs 
$b$ and $a'$, $a'$ and $c'$, and $c'$ and $b'$
share one edge). 
For more details on circular orderings, see \cite{sem}.

Now, let $(a_0,a_1,\dots,a_{n-2},a_{n-1})$ be a circular ordering for a binary $X-$tree
$T=(V,E)$,
and put
$$
\cl: = \{a_0 a_i: i=1, \ldots, n-1\} \cup \{a_{i-1} a_i: i= 2, \ldots, n-1\}.
$$
Then, 
$|\cl| = |T|=2n-3$ holds and  $\cl$ is a triplet as well as a pointed  $a_0-$cover of $T$, and it is a  well-known fact that
$\cl$ lassos the edge weights of  $T$  \big(see Proposition 2.3 of \cite{bar} or \cite{cha} for
the case where $\omega(e)=1$ for all $e\in E$\big).

\begin{figure}[ht]
\begin{center}
\resizebox{8cm}{!}{
\includegraphics{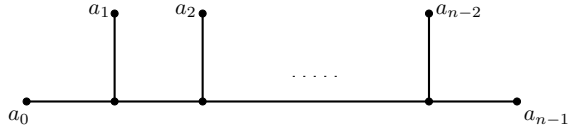}
}
\caption{A caterpillar tree 
for which $(a_0,a_1,\dots,a_{n-2},a_{n-1})$ is a circular ordering.
}
\label{caterfig}
\end{center}
\end{figure}

Further, if $T$ is a caterpillar tree  with the two cherries
$a_0,a_1$ and $a_{n-2}, a_{n-1}$  and
$(a_0,a_1,\dots,a_{n-2},a_{n-1})$ is a
circular ordering for $T$ relative to the planar embedding of $T$ that is indicated in
Fig.~\ref{caterfig},
then the union $\cl$ of the sets $\{a_0x:x\in X, x\neq a_0\}$ and $\{a_{n-1}x:x\in X,  x\neq a_{n-1}\}$ is a
triplet as well as a pointed
$a_0-$cover of $T$ for which $|\cl| = |E|=2n-3$ holds.

Theorem \ref{cbU} implies the following result.
\begin{proposition}\label{triple}
Every triplet cover $\cl$ of a binary $X-$tree $T$ lassos the edge weights for $T$. Furthermore, $(T,\omega)\equiv (T',\omega')$ must hold for every proper edge weighting $\omega$ of $T$
and every pair $(T',\omega')$ that consists of an $X-$tree $T'$ and  an edge weighting $\omega'$ of $T'$ such that $\cl$ is also a triplet cover of $T'$ and $(T,\omega)\eb (T',\omega')$ holds.
\end{proposition}
\pf Choose any 
$T-$cherry $a,b$
and note that, with $U=U_{ab}:=V-\{a,b\}$ and all the 
notations and
conventions introduced in the context of Theorem \ref{cbU}, the subset $\cl_U$ of $X_U\choose 2$
is a triplet cover of $X_U$ whenever $\cl$ is
a triplet cover of $T$. Thus, we may assume that, by induction, $\cl_U$ satisfies our claims for $T_U$. Moreover, every triplet cover $\cl$ of $T$ must contain the cord $ab$ and there must exist some $c\in X-\{a,b\}$ with $ac,bc\in \cl$. So, all the assertions $({\bf U1}),({\bf U2}),$ and
$({\bf U3-a})$ must hold for $\cl$, implying that
$\cl$ is indeed an edge-weight lasso for $T$.

Furthermore, given any proper edge weighting $\omega$ of $T$, the pair $a,b$ must form a 
$T'-$cherry in every $X-$tree $T'$ 
for which $\cl$ is also a triplet cover of $T'$ and
an edge weighting $\omega'$ of $T'$ with $(T,\omega)\eb (T',\omega')$ exists.  Indeed, it is obvious that, given any triple $a',b',c'$ of distinct leaves with $a'b',a'c',b'c'\in \cl$, the interior vertex ${\rm med}_T(a',b',c')$ is adjacent to $a'$ in $T$ if and only if we have
$$
D_\omega(a'a'|b'c')=\min(D_\omega(a'a'|yz): y,z\in X-\{a'\}; a'y,a'z,yz
\in \cl),
$$
(note that, by definition (cf. \ref{D(|)}),  $D_\omega(aa|bc)=D_\omega(a,b)+D_\omega(a,c)-D_\omega(b,c)$ holds for all $a,b,c\in X$).
 Thus, if ${\rm med}_T(a',b',c')$ is adjacent to $a'$ for some $a',b',c'\in X$ as above, this must also be true for the vertex
${\rm med}_{T'}(a',b',c')$ in $T'$, provided that $(T,\omega)\eb (T',\omega')$ holds. In particular, the pair $a,b$ must form a 
$T'-$cherry if it forms a$T-$cherry and 
$(T,\omega)\eb (T',\omega')$
holds. It follows that
$(T_U,\omega_U)\eu (T'_U,\omega'_U)$ must also hold and, by induction, therefore  $(T_U, \omega_U) \equiv (T'_U,  \omega'_U)$ also holds, which easily implies our claim
 $(T,\omega)\equiv(T',\omega')$.
\epf

Regarding pointed covers, even a stronger result holds:
\begin{theorem}\label{pointed}
If a subset $\cl$ of $\ch$ is a pointed cover of a binary $X-$tree $T
$, then $\cl$ is an $s-$lasso and, hence, a strong lasso for $T$.
More specifically, if $\cl$ is a pointed $x-$cover of $T$ for some $x\in X$, then there exists an 
``$x-$shelling''
$a_1 b_1, a_2 b_2, \ldots a_m b_m$ of $\ch-\cl$, i.e.,
a shelling such that, for every $\mu = 1,\dots,m$, one of the two
pivots $x_\mu, y_\mu$ for  $a_\mu b_\mu$ can be chosen to coincide
with $x$.
\end{theorem}
\pf  Clearly, we may assume, without loss of generality, that $n\ge 5$ holds.
Consider a (necessarily proper) 
$T-$cherry $a,b$ 
not containing $x$. As above, we put $U=U_{ab}:=V-\{a,b\}$ and use all the 
notations and
conventions introduced in the context of Theorem \ref{cbU}.

\begin{figure}[ht]
\begin{center}
\resizebox{4cm}{!}{
\includegraphics{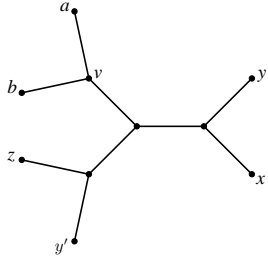}
}
\caption{One of the various configurations that can occur for $T$ if one adds the vertices $y,y',z$ to the three vertices $a,b,x$
(see text for details).}
\label{latex_fig5}
\end{center}
\end{figure}

First note that, if we have any two distinct elements $y,z\in X-\{x,a,b\}$,
the tree $T|_{\{a,b,x,y\}}$ obtained from $T$ by restriction to $\{a,b,x,y\}$ is always a quartet tree of type $ab\|xy$. Moreover, the two trees
$T|_{\{a,x,y,z\}}$ and $T|_{\{b,x,y,z\}}$ obtained from $T$ by restriction to $\{a,x,y,z\}$ and $\{b,x,y,z\}$ are, respectively, quartet trees of type $xy\|az$ and
$xy\|bz$ in case the tree $T_U|_{\{v,x,y,z\}}$
 obtained from $T_U$ by restriction to $\{v,x,y,z\}$ is a quartet tree of type $xy\|vz$,
and these two trees are, respectively, quartet trees of type $xa\|yz$ and
$xb\|yz$ in case $T_U|_{\{v,x,y,z\}}$ is a quartet tree of type $xv\|yz$ -- see Fig. \ref{latex_fig5} for an illustration.

\medskip

Now assume that  $\cl \subseteq \ch$ is a pointed $x-$cover of $T$
for some leaf $x\in X$.  It is obvious that $\cl_U$ is an pointed $x-$cover of $T_U$. So, by induction, there must exist a shelling $a_1b_1, a_2b_2, \dots, a_mb_m$ of ${X_U\choose 2}-\cl_U$ such that, for every $\mu\in \{1,2,\dots,m\}$, there exists some element $y_\mu$ in $X_U-\{x, a_\mu,b_\mu\}$ for
which  the tree $T_U|_{\{a_\mu,b_\mu,x,y_\mu\}}$ obtained from $T_U$ by restriction to
$\{a_\mu,b_\mu,x,y_\mu\}$ is a quartet tree of type $a_\mu x\|y_\mu b_\mu$ and
all cords in ${\{a_\mu,b_\mu,x,y_\mu\}}\choose{2}$ except $a_\mu b_\mu$ are contained in $\cl_{U,\mu}:=\cl_U \cup \{a_{\mu'}b_{\mu'}: \mu'\in  \{1,2,\dots,\mu-1\}\}$.
So, to produce an $x-$shelling of  $\ch-\cl$, we may first take 
all cords in $\ch-\cl$ of the form $ay$ with $by \in \cl$ and use
$b$ as their second pivot.

 Noting that the tree
 $T|_{\{a,b,x,y\}}$ is a quartet tree of type $ab\|xy$ and that all cords in $\{a,b,x,y\}\choose 2$ except for $ay$, are contained in $\cl$, we can take these cords in any order.
 
 Then, we take all  cords in $\ch-\cl$ of the form $by$ with
$ay \in \cl$ in any order and use $a$ as their second pivot,
 which works for the same reason. Then, for each $\mu=1,2,\dots,m$ with $v\not\in \{a_\mu,b_\mu,y_\mu\}$, we take the cord $a_\mu b_\mu$ and use $y_\mu$ as its second pivot. In case $y_\mu =v$, we take the cord $a_\mu b_\mu$ and  use $a$ or $b$ as its second pivot. In case  $a_\mu =v$, we take the cord $ab_\mu$ and  use $y_\mu$ as the second pivot and then add the cord $bb_\mu$, taking $a$ as the second pivot. And 
 finally, 
 in case $b_\mu =v$, we switch
$a_\mu$ and $b_\mu$
and, otherwise,
proceed as above.
Now, a 
simple inductive argument
shows that this defines an $x-$shelling of $\ch-\cl$ as required.\epf\

Our earlier Example \ref{shelex} illustrates Theorem \ref{pointed}, as 
$\{ab, bc, cd, de, ea, ad, ac\}$
is obviously a pointed $a-$cover of $T_5$.

Let us finally return to the general setting of 
 (not necessarily binary) $X-$trees
and consider subsets $\cl$ of $\ch$ that are of the form $$\cl=A\vee B:=\big\{\{a,b\}: a\in A, b\in B\big\}$$ for some 
$X-$split $A,B$.
For example (cf.  Fig.~\ref{figure_quartet}), 
the topological lasso $\cl_4$ for $T_4$ is of this form as it coincides with $\{a,d\}\vee \{b,c\}$.

It follows immediately from our definitions that, given any $2$-subset $\cy=\{x,x'\}$ of $X$ with $x\in A$ and $x'\in B$ and any subset $Y$ of $X-\cy$, the restriction
$\Gamma(\cl,\cy)|_Y$ of the graph $\Gamma(\cl,\cy)$ introduced in Section \ref{facts} is the complete bipartite graph with vertex set $Y$ whose edge set is $(A\cap Y)\vee(B\cap Y)$. Thus, it is connected if and only if one has either $|Y|=1$ or neither $A\cap Y$ nor $B\cap Y$ are empty. And it is also obvious that, whenever $\cl=A\vee B$ is a $t-$cover of an $X-$tree $T$, one must have $A \cap \cy\neq \emptyset \neq B \cap \cy$ for every $2$-subset $\cy$ of $X$ whose elements form a
$T-$cherry.

$\cl$ can therefore only be a $t-$cover of an $X-$tree $T$ if there exist
no three distinct leaves
$a,b,c$ of $T$ with $v_a = v_b=v_c$.
 Thus, Theorem \ref{top} implies the following characterization of topological lassos $\cl$ that are of the form $\cl=A\vee B$ for some 
 $X-$split $A,B$:

\begin{theorem}
\label{vee}
Given an $X-$tree $T$ and  
an $X-$split
$A,B$ of $X$, the following four assertions are equivalent:
\begin{itemize}
\item[{\rm (i)}] The subset $A\vee B$ of $\ch$ is  a topological lasso for $T$.
\item[{\rm (ii)}] $A\vee B$ is  a $t-$cover of $T$.
\item[{\rm (iii)}] $A \cap \cy\neq \emptyset \neq B \cap \cy$ holds for every $2$-subset $\cy$ of $X$ whose elements form a 

$T-$cherry.
\item[{\rm (iv)}] The bipartition $A,B$ of  $X$ is incompatible with every non-trivial virtual $T-$split.
\end{itemize}
\end{theorem}
\pf It is obvious from the definitions and previously recorded facts that ``(i) $\Ra$  (ii) $\Ra$  (iii)''  holds. The implication ``(iii) $\Ra$  (iv)'' holds because any subset $A'$ of $X$ for which $A',X-A'$ is a non-trivial virtual $T-$split must contain two elements that form a 
$T-$cherry.

And ``(iv) $\Ra$  (i)'' holds in view of Theorem \ref{top}: First observe 
that there exists always a $2$-subset $\cy$ of $X$ whose elements form a proper $T-$cherry\footnote{To see this, just take any leaf $a$ of maximal distance to some other arbitrary 
vertex $u$. Then, $a$ must be part of a cherry $a,b$  whose elements 
$a,b$ must be adjacent to an interior vertex $v=v_a=v_b$ that is 
incident with only one interior edge (by the maximal distance to $u$ assumption) and with no other pendant edge (by the condition imposed in (iv)).  So, $a,b$ must necessarily be a 
proper cherry.}.
Note next that, for any such $2$-subset $\cy$,
there must exist $x\in A$ and $x'\in B$ with $\cy=\{x,x'\}$ (as $A,B$ is supposed to be incompatible with every non-trivial virtual $T-$split and, hence, in particular with the $T-$split $\cy,X-\cy$), and that (for the same reason) $A\cap Y\neq \emptyset \neq B\cap Y$ must hold for every subset $Y$ of $X-\cy$ for which $Y,X-Y$ is a non-trivial virtual $T-$split implying
that $\Gamma(\cl,\cy)|_Y$ must be connected (as required in Theorem \ref{top}).
\epf

\section{Remarks and questions}
\label{remarks}
Our results  raise further questions concerning the properties of different types of lassos:
\begin{itemize}
\item[Q1.] Does there exist a triplet cover of a {\em binary} tree that is not a strong lasso?
\item[Q2.]  Can we characterize those covers of a binary $X-$tree $T$ that are a
tight  edge-weight or strong lasso of $T$?
\end{itemize}

Regarding the second question, two necessary conditions for a cover of a binary$X-$tree $T$ of cardinality $2n-3$ to lasso the edge weights of $T$ are as follows:
\begin{itemize}
\item
For each subset $Y$ of $X$ of cardinality $m$, the cardinality of $\cl\cap \binom{Y}{2}$ cannot exceed $2m-3$ (to avoid over-determination at $Y$).
\item

If  $ab,bc,cd,da\in \cl$ holds for some four
leaves $a,b,c,d\in X$, then no edge of $T$ can
separate $a,c$ from $b,d$ (as this would imply that $D(a,b)+D(c,d) =
D(b,c)+D(d,a)$
  would hold).
\end{itemize}

It may also be of interest to investigate further the properties of weak lassos.
  Note that if $T$ is a binary $X-$tree,
then $\cl$ is a weak lasso for $T$ if and only if $\cl$ is a topological lasso for  $T$; however, for non-binary trees,
these are quite different concepts.   For example, any subset of $\binom{X}{2}$ (including the empty set) is a weak lasso for the star tree  
$T^*$ with leaf set $X$
since any $X-$tree is a resolution of that tree, but it requires all of $\binom{X}{2}$ to lasso the shape of $T^*$.

Finally, we can view a triplet cover as a subset of $\ch$
that contains all the cords ``induced'' by a sufficiently large
collection of subtrees each of which has three leaves.
Thus, 
it is mathematically natural, and relevant to phylogenetic analysis (supertree reconstruction),
to study the lasso properties of subsets of $\binom{X}{2}$ that are induced by collections of phylogenetic trees with three or more leaves.
More precisely, given an $X-$tree $T$ and a collection
$\cP = \{X_1 , \dots , X_k\}$ of subsets of $X$, let
 $$\cl_{\cP} := \bigcup_{i=1}^k\binom{X_i}{2}.$$

It would be of interest to  determine conditions on  $\cP$ in order for $\cl_\cP$ to lasso $T$, at least in case $T$ is binary. The quartet case
(where all sets in $\cP$ are $4$-subsets of $X$) is an obvious
candidate for analysis, in view of a range of combinatorial results
from \cite{boc}, \cite{dre}, \cite{Dre10}, \cite{gru} and \cite{ste}.\\

\noindent{\bf Acknowledgements}
A.D.  thanks the CAS and the MPG for financial support;   K.T.H. was partially supported by the Engineering and Physical Sciences Sciences Research Council [grant number EP/D068800/1]. Also she
would like to thank the Department of Mathematics and Statistics, University of Canterbury, New Zealand, for hosting her during part
of this work. M.S. thanks the Royal Society of NZ under its Marsden Fund and James Cook Fellowship scheme.  We also thank the two anonymous reviewers for several helpful suggestions. 
The final publication is available at springerlink.com
under:

\noindent http://www.springerlink.com/content/ppk5341254402164/. 

\noindent (DOI: 10.1007/s00285-011-0450-4).


\begin{thebibliography}{}

\bibitem{att} Atteson K (1999) The performance of neighbor-joining methods of phylogenetic reconstruction. Algorithmica 25: 251--278.
\bibitem{bar}  Barth{\'e}lemy JP, Gu{\'e}oche A (1991) Trees and Proximity Representations. John Wiley and Sons Ltd.
\bibitem{boc} B{\"o}cker S, Dress AWM, Steel M (1999) Patching up $X-$trees. Ann Combin 3: 1--12.
\bibitem{cha} Chaiken S, Dewdney AK, Slater PJ (1983) An optimal diagonal tree code. SIAM J Alg Disc Math 4(1): 42--49.
\bibitem{col} Colonius, H. and Schulze, H. H. (1981). Tree structures for proximity data. British Journal of Mathematical and Statistical
Psychology, 34: 167--180.
\bibitem{dre} Dress A, Erd{\"o}s PL (2003) $X-$trees and weighted quartet systems. Ann Combin 7: 155--169.
\bibitem{dre2} Dress A, Steel M (2009) A Hall-type theorem for triplet set systems based on medians in trees. Appl Math Lett 22: 1789--1792.
\bibitem{dres2} Dress, A., Huber, K. Steel, M. Lassoing a phylogenetic tree (II), manuscript in preparation.
\bibitem{Dre10} Dress A, Huber K, Koolen J, Moulton V, Spillner A (2011) Basic Phylogenetic Combinatorics. Cambridge University Press (in press).
\bibitem{far} Farach M, Kannan S, Warnow T (1995) A robust model for finding optimal evolutionary trees. Algorithmica 13: 155--179.
\bibitem{fel} Felsenstein J (2004) Inferring Phylogenies. Sinauer Associates, Sunderland, MA.
\bibitem{gue2} Gu{\'e}noche A, Leclerc B (2001) The triangles method to build $X-$trees from incomplete distance matrices. ROADEF'99 (Autrans) RAIRO Oper Res 35(2): 283--300.
\bibitem{gue} Gu{\'e}noche A, Leclerc B, Markarenkov V (2004) On the extension a partial metric to a tree metric. Discr Appl Math 276: 229--248.
\bibitem{gru} Gr{\"u}newald S, Huber KT, Moulton V, Semple C (2008) Encoding phylogenetic trees in terms of weighted quartets. J Math Biol 56: 465--477.
\bibitem{phil} Philippe H, Snell E, Bapteste E, Lopez P, Holland P, Casane D (2004) Phylogenomics of eukaryotes: Impact of missing data on large alignments. Mol Biol Evol 21: 1740--1752.
\bibitem{sem} Semple C, Steel M (2003) Phylogenetics. Oxford University Press.
\bibitem{ste} Steel MA (1992) The complexity of reconstructing trees from qualitative characters and subtrees. J Classif 9: 91--116.
\bibitem{will} Willson SJ (2004) Computing rooted supertrees using distances. Bull Math Biol 66(6): 1755--1783.




\end{thebibliography}
\end{document}